\documentclass[usenatbib]{mn2e}
\usepackage{times,graphics,amssymb,epsfig,subfigure,amsmath}
\usepackage{cite,aas_macros,url}
\usepackage{hyperref}
\usepackage{color}

\newcommand{\de}{{\rm d}}
\newcommand{\bea}{\begin{eqnarray}}
\newcommand{\eea}{\end{eqnarray}}
\newcommand{\f}{\frac}
\defcitealias{2008MNRAS.385..783S}{Paper~I}
\defcitealias{2010MNRAS.402.2778S}{Paper~II}
     
\title[Cosmic ray driven outflows]
{Efficient cold outflows driven by cosmic rays in high redshift galaxies
and their global effects on the IGM}
\author[Samui, Subramanian \& Srianand] 
{Saumyadip Samui$^{1}$\thanks{E-mail: saumyadip.physics@presiuniv.ac.in},
	Kandaswamy Subramanian$^2$\thanks{E-mail: kandu@iucaa.in},
        Raghunathan Srianand$^2$\thanks{E-mail: anand@iucaa.in}\\
	$^1$Presidency University, 86/1 College Street, Kolkata 700073, India\\
        $^2$IUCAA, Post Bag 4, Ganeshkhind, Pune 411 007, India.}

\date{Accepted XXX. Received YYY; in original form ZZZ}
\pubyear{2017}

\begin{document}
\label{firstpage}
\pagerange{\pageref{firstpage}--\pageref{lastpage}}
\maketitle 
\begin{abstract}
We present semi-analytical models of galactic outflows in high redshift
galaxies driven by both hot thermal gas and non-thermal cosmic rays.
Thermal pressure alone may not sustain a large scale
outflow in low mass galaxies (i.e $M\sim 10^8$~M$_\odot$), in the presence
of supernovae (SNe) feedback with large mass loading.
We show that inclusion of cosmic ray pressure allows outflow solutions
even in these galaxies.
In massive galaxies for the same energy efficiency, cosmic ray
driven
winds can propagate to larger
distances compared to pure thermally driven winds. On an average gas
in the cosmic ray driven winds has a lower temperature
which could aid detecting it
through absorption lines in the spectra of background sources.
Using our constrained semi-analytical models of galaxy formation
(that explains the observed UV luminosity functions of galaxies) we study
the influence of cosmic ray driven winds on the properties of the
intergalactic medium (IGM) at different redshifts. In particular,
we study the volume filling factor, average metallicity, cosmic ray
and magnetic field energy densities for models invoking
atomic cooled and molecular cooled halos. We show that the
cosmic rays in the IGM could have enough energy that can be transferred to the
thermal gas in presence of magnetic fields to influence the thermal
history of the intergalactic medium. The significant volume filling
and resulting strength of IGM magnetic fields can also account for
recent $\gamma$-ray observations of blazars.
\end{abstract}

 
\section{Introduction} 
Energy injection by supernova explosions can lead to strong outflows from
star forming galaxies.
These supernovae (SNe) driven galactic outflows could eject a significant 
fraction of the interstellar medium (ISM) along with metals
into their circum-galactic medium (CGM) and the general 
intergalactic medium (IGM). 
Feedback due to starburst driven outflows are also invoked in galaxy formation
models to get correct shape of the galaxy luminosity functions.
It is well demonstrated that a high fraction (40-60\%) of high-$z$
star forming galaxies show signatures of bi-conical outflows
in the form of high velocity (i.e. 100-300~km/s) blue shifted absorption
in their spectrum \citep{2012ApJ...760..127M,2014ApJ...794..156R}.
Further, quasar spectra reveal presence of metals in the form of
absorption by ions like
C~{\sc iv},
O~{\sc vi} etc in the low density intergalactic medium far away from the
star forming galaxies \citep{1996AJ....112..335S,2002ApJ...578...43C,
2006MNRAS.371L..78R,2013ApJ...763...37C,2013MNRAS.435.1198D,
2012MNRAS.421..446M,2016MNRAS.463.2690D}.
Moreover, the CGM has been mapped out to a
projected distance of few 100~kpc from the host galaxies using
galaxy-quasar pairs \citep[see for example][]{2013ApJ...776..114N}.
This has revealed the presence of metals in
the cold gas 
at $10^4$~K (traced by Mg~{\sc ii} and/or Ca~{\sc ii})
as well as in highly ionised warm gas
\citep{2010ApJ...714.1521C,2011ApJ...740...91P,2013ApJ...770...41M,
2014ApJ...796..140P} in the CGM.
As metals are only produced in stars, they have to
be transported into the general IGM or the CGM by large scale galactic outflows.

SNe driven galactic outflows have been extensively studied using both 
hydrodynamical simulations 
\citep{2005MNRAS.364..552S,2006MNRAS.371.1125S,2008MNRAS.391..110D,
2011MNRAS.415...11D,2012MNRAS.423.1726S}
and semi-analytical models
\citep{1988RvMP...60....1O,2001ApJ...555...92M,2002ApJ...574..590S,
2003ApJ...588...18F,2008MNRAS.385..783S,2013ApJ...763...17S}.
In a previous paper we studied thermally driven outflows from high redshift
galaxies and their global effects on the IGM, through a semi-analytic model 
\citep[][hereafter Paper I]{2008MNRAS.385..783S}.
Our outflow model
was similar to models of stellar wind blown bubbles \citep{1985Natur.317...44C}. Here, the free wind from
a galaxy converts its kinetic energy to thermal energy through an inner shock
and feeds a hot bubble that expands into the halo medium (or CGM) and subsequently the 
general IGM (see Fig.~\ref{wind_profile}  below
for the entire outflow structure). We showed that outflows can efficiently
pollute the CGM and the IGM with metals. However, their detectability
in the form of absorption was not clear due to the high temperature and
low density of
the outflowing gas. Nevertheless, a significant volume of the IGM could be
filled with metals due to outflows dominated by low mass galaxies
(with dark matter halo mass less than $10^{10}$~M$_\odot$).

Strong shocks created by exploding SNe also accelerate
highly relativistic charged particles,
usually referred to as
cosmic rays (CRs). 
Cosmic rays gyrate along the magnetic
field lines and transfer energy and momentum to the thermal gas via
Alfv\'en waves
\citep{1968ApJ...152..987W,1971ApJ...163..503W,1969ApJ...156..445K,
1971ApL.....8..189K}.
In \citet[][hereafter Paper II]{2010MNRAS.402.2778S},
we explored the possible influence of
cosmic rays  generated in the SNe shocks on free wind solutions 
assuming a constant rate of star formation.
The addition of energy and momentum due to cosmic rays
helps to drive the free wind at the base of the outflow.
It is important especially in low mass galaxies where the gas could lose 
its thermal energy rapidly due to radiative cooling.
Further, such cosmic rays can be reaccelerated at the inner-shock of outflows
and add pressure to the hot bubble that drives an outer shock (see Fig.~\ref{wind_profile}).
Also, the cosmic rays in the hot bubble do not 
cool radiatively and their pressure decreases less steeply with 
adiabatic expansion compared to thermal gas due to the differences
in the adiabatic indices. Therefore, it is important to see how their
presence can influence the outflow dynamics.
Moreover, there is an increasing interest in exploring the consequences 
of incorporating CRs in simulations of outflows, which all point to their 
usefulness in driving outflows
(\citealp{2014MNRAS.437.3312S,2016ApJ...816L..19G,
2016ApJ...824L..30P,2017ApJ...834..208R,2017MNRAS.465.4500P};\citealp[and also see the review
by][section 3.4 which emphasizes the role of CRs]{2017ARA&A..55...59N}).
Thus incorporating the effect of CRs in our semi-analytical outflow
models to examine their global consequences for the IGM is also timely.

Further, our cosmic ray free-wind solutions showed that the mass
loading factor $\eta_w$, 
(ratio of mass loss rate due to wind to star formation rate),
goes as $v_c^{-2}$, where $v_c$ is the circular velocity of the halo.
This implies a strong negative feedback on star formation 
in low mass galaxies \citep{2010MNRAS.402.2778S}.
It is then not obvious that such a reduced star formation
can indeed drive and sustain strong outflows
that go beyond the virial radius
in low mass galaxies. 
Our earlier thermally driven outflow models \citepalias{2008MNRAS.385..783S} also 
did not take into account this negative feedback in star formation.
Thus, it is important to revisit outflow models considering not only 
the effects of
cosmic ray driving but also incorporating the negative feedback on
star formation implied by such outflows. 
An important ingredient in outflow models is the star formation
rate in a forming galaxy.
In a series of papers \citep{2007MNRAS.377..285S,2009MNRAS.398.2061S,2009NewA...14..591S,2013MNRAS.429.2333J,2014NewA...30...89S}, we have built up a semi-analytical picture
of galaxy formation which reproduces a variety of observations of the high
redshift universe. These include UV \& Lyman-$\alpha$ luminosity
functions, clustering of Lyman-break
galaxies and Lyman-$\alpha$ emitters, the correlation between star formation
rate, stellar mass and halo mass of galaxies. In particular, we use here
the models of \citet{2014NewA...30...89S} which incorporated
supernova feedback in determining the star formation rate in the high redshift
galaxies.

There are several additional reasons to study such outflow models.
Note that cosmic rays always remain coupled to magnetic fields which are
themselves coupled to the thermal gas.
Therefore, when winds transport metals into the
IGM, they will also carry along magnetic fields and cosmic rays. Thus outflows
spread magnetic fields that are being generated/amplified in galaxies to the
intergalactic medium. Recent $\gamma$-ray observations of distant blazars
have even suggested a lower limit to the magnetic fields in void
regions at a level of $10^{-16}$G if correlated on Mpc scale
\citep{2010Sci...328...73N}.  \citet{2006MNRAS.370..319B}
have studied possibility of IGM magnetisation by galactic outflows.
It is interesting to
reconsider this issue with our new improved models of outflows and estimate
the strength and volume filling factor of
magnetic fields that can be seeded by these outflows into the IGM. 

Further, if
sufficient magnetic fields and cosmic rays are present in the IGM, cosmic
rays will still be able to transfer their residual energy to the thermal gas via
Alfv\'en waves. 
Pockets of over pressured cosmic ray regions can also transfer energy
to the IGM via adiabatic heating as they expand and come
into pressure equilibrium with their surroundings.
If the remaining cosmic ray energy (that is after driving
the outflows) is comparable to the thermal energy of the IGM it can
potentially influence the thermal history of the IGM. This is interesting
as observations have suggested a rise in IGM temperature at $z\sim 3$ from
combining simulations and quasar spectrum 
\citep{2000ApJ...534...41R,2011MNRAS.410.1096B,2014MNRAS.441.1916B}.
Some authors have attributed this to Helium reionization
\citep{2000MNRAS.318..817S,2008ApJ...682...14F,2009MNRAS.395..736B}.
However, cosmic rays could also play an
important role in heating the IGM \citep{2005ICRC....9..215S},
a possibility that we also explore in this work.

The paper is organised as follows.
In section~\ref{sec_outflow}, we outline our outflow models. Magnetic
field evolution is described in section~\ref{sec_mag_field}. The
characteristics of individual outflows is studied 
in section~\ref{sec_wind_profile}. The
global impact of outflows is discussed in section~\ref{sec_global}. 
The implications of such outflows are addressed in section~\ref{sec_impl}.
Finally
section~\ref{sec_dc} presents a discussion of our results and the conclusions. 
Through out this work
we assume cosmological parameters as suggested by Planck results,
namely, a flat $\Lambda$CDM cosmology with $\Omega_\Lambda = 0.70$,
$\Omega_b=0.044$, $n_s=0.96$, $H_0=71$~km/s/Mpc and $\sigma_8=0.8$
\citep{2016A&A...594A..13P}.

\section{Outflow dynamics}
\label{sec_outflow}

\begin{figure}
\centerline{
\epsfig{figure=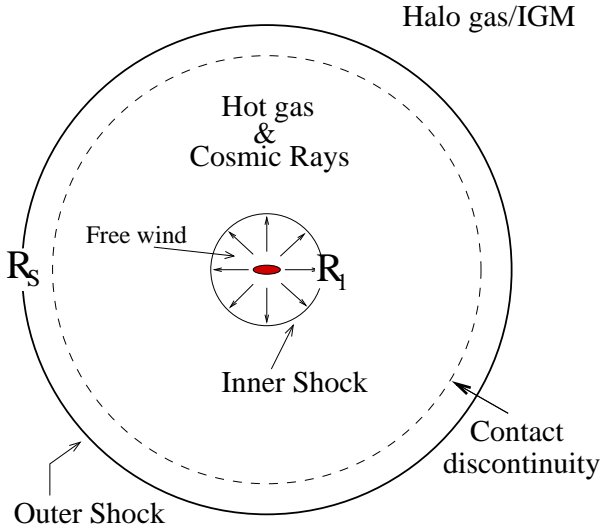,width=8.0cm,angle=0.}}
\caption[]{Wind profile for thermally and cosmic ray driven winds.
}
\label{wind_profile}
\end{figure}

The structural properties of the outflow remain by and large the
same as in \citetalias{2008MNRAS.385..783S}.
However, we improve this model by incorporating
the extra pressure due to the non-thermal cosmic ray particles
in the outflow dynamics.
We assume a spherically symmetric thin shell model of outflows from galaxies
\citep{1977ApJ...218..377W,1988RvMP...60....1O}.
The coherent explosions of SNe in a galaxy produce a bubble of hot gas
that expands and expels gas from ISM \citep[][]{2014MNRAS.443.3463S}. The hot bubble expands as
the `free wind' coming from subsequent SNe explosions produces a shock
and converts its kinetic energy to thermal energy.
The free wind consists of thermal as well as non-thermal particles (CRs)
that are produced at the SNe terminal shocks.
Such `free wind' solutions with both thermal and cosmic ray components
have been investigated in \citetalias{2010MNRAS.402.2778S}
\citep[Also see][]{1975ApJ...196..107I,1991A&A...245...79B}.
The asymptotic velocity of such a free wind is closely related to the
circular velocity of the halo. If the halo gas is of sufficient density
the free wind produces a shock and
converts its kinetic energy to the thermal energy that feeds the hot
bubble. We call this as `inner shock' of the outflow. At this inner shock
cosmic rays can also be accelerated through diffusive shock acceleration
mechanism with efficiency as high as 50\% \citep{2003ICRC....4.2039K,
2005ApJ...620...44K}
in transferring the kinetic energy of the wind to CR particles.
The thermal
pressure of the gas as well as the cosmic ray pressure in the hot bubble
drives an `outer shock' and sweeps up CGM/IGM material. The swept up
material by the outer shock remains confined to a thin shell region
and separated from the shocked free wind material by contact discontinuity.
In fact, self-similar solution of cosmic ray and thermal gas driven
outflow shows that most of the mass of the hot bubble is also expected
to be concentrated at the contact discontinuity \citep{2015MNRAS.447.2224B}.
See Fig.~\ref{wind_profile} for the schematic diagram 
of our outflow models.

The evolution of such spherically symmetric outflows is 
governed by the following equations in the presence of cosmic rays and
the hot gas
\citep[cf.][]{1977ApJ...218..377W,1988RvMP...60....1O,1993ApJ...417...54T},
\bea
\f{\de ^2 R_s}{\de t^2} &=& \f{4\pi R_s^2 \left[P_b +P_c -P_0\right]}{m_s(R_s)}
- \f {\dot m_s(R_s)\left[\dot{R}_s - v_0(R_s)\right] }{m_s(R_s)}
\nonumber \\
 & & \:\:\:\:\:\:\:- \f {G M(R_s)}{R_s^2}, 
 \label {eqnradius} 
\eea
\begin{equation}
\f{\de}{\de t}\left[m_s(R_s)\right] =
 \epsilon 4 \pi R_s^2 \rho_B(R_s) \left[\dot{R}_s - v_0(R_s)\right].
 \label {eqnmaass} 
\end{equation}
Here, the position of the outer shock is $R_s$, $m_s$ is the gas
mass of the thin shell and $P_b$ and $P_c$ are the thermal and
cosmic ray pressures in the hot bubble. Further, $\rho_B$ and $P_0$ are respectively
the baryon density and the pressure
of the outside medium assuming virial temperature for halo gas or $10^4$~K
for the IGM; 
$v_0$ is the ambient velocity field taken from \citet{2001ApJ...556..619F}.

The first term in the right hand side of Eq.~\ref{eqnradius} represents
the net pressure i.e. thermal plus cosmic ray pressure minus the outside
pressure, that drives the outer shock. The second term is due to the mass
loading of the shell from the outside medium causing the deceleration of the shock
and we assume a fraction $\epsilon$ of all swept up mass remains in the
thin shell (i.e. Eq.~\ref{eqnmaass}) and rest is mixed with the hot bubble
due to evaporation/fragmentation
(the mass in the hot bubble is negligible compare to the swept up
mass).
The 3rd term in Eq.~\ref{eqnradius} is the gravitational attraction due
to the dark matter \citep[a NFW profile is assumed with concentration parameter
$c=4.8$,][]{1997ApJ...490..493N,2001ApJ...555...92M}
 as well as baryonic matter inside the radius $R_s$. For the initial baryonic
density profile inside the halo we assume a beta model \citep{1998ApJ...497..555M}
and assume that a fraction, $f_h=0.1$ of total the total gas remains in the halo.

The thermal pressure $P_b$ can be obtained from the thermal energy of the bubble, $E_b$, by  
\begin{equation}
P_b = \frac{E_b}{ 2\pi (R_s^3-R_1^3)},
\label{pressure_energy}
\end{equation}
where we assume that the adiabatic index of the thermal gas $\gamma = 5/3$. 
Here, $R_1$ is the position of the inner shock of the outflow.
Similarly, the non-thermal cosmic ray pressure $P_c$ and the cosmic ray
energy in the bubble, $E_c$, are related by, 
\begin{equation}
P_c = \frac{U_c}{3}=\frac{E_c}{ 4\pi (R_s^3-R_1^3)},
\label{pressure_energy_cr}
\end{equation}
with the assumption that the adiabatic index of the CR component $\gamma_{CR} = 4/3$ (considering them as highly relativistic particles). Here, $U_c$ is the energy
density of the cosmic ray particles.
The evolution of the thermal energy of the bubble is given by,
\begin{equation}
\f {\de E_b}{\de t} = L(t) - \Lambda (t,T,Z) 
   - 4\pi \left[R_s^2 \dot{R}_s - R_1^2 \dot{R}_1\right] P_b.
    \label {eqnenergy}
\end{equation}
The `free wind' carries some fraction of total SNe energy in the host galaxy
which is fed into the bubble as thermal energy, $L(t)$, at the inner shock.
The bubble can
lose its thermal energy due to $P dV$ work as it expands and radiative cooling.
For cooling rate i.e. $\Lambda (t,T,Z)$, we consider Compton drag against the CMBR, bremsstrahlung
and recombination line cooling that depends on temperature $T$ and
metallicity, $Z$ of the gas. The gas density that determines the temperature of
the hot bubble is taken to be the average density of the bubble. Note that the
bubble is filled by the gas coming from the free wind and a fraction
of the swept up mass. The free wind mass is higher for low mass galaxies
owing to higher mass loading factor, 
$\eta_w$, that leads to higher radiative cooling of the bubble.
Further, the metallicity of the
bubble gas is self-consistently determined as described in the Appendix~A
of \citetalias{2008MNRAS.385..783S}.

Similarly, the evolution of the cosmic ray energy in the bubble 
is governed by
\begin{equation}
\f {\de E_c}{\de t} = L_c(t)  - 4\pi \left[R_s^2 \dot{R}_s - R_1^2 \dot{R}_1\right] P_c.
    \label {eqnenergycr}
\end{equation}
Here, $L_c$ is the total energy of the cosmic ray component per unit time
injected into the bubble,
at the inner shock. Note we consider, at the inner shock,
the cosmic rays are being accelerated through diffusive shock acceleration
with efficiency as high as 50\% \citep{2003ICRC....4.2039K,2005ApJ...620...44K}
in converting
the kinetic energy of the shock front into cosmic rays. Thus, if initially
we have 10\% of total supernova energy available in the free wind, at the
inner shock half of it will be available as thermal energy and rest half
will be converted into cosmic ray energy. It is interesting to note that
cosmic rays lose energy only due to adiabatic expansion that is solely
determined from the outflow dynamics. But the thermal energy is lost by
additional cooling processes and can be significant fraction of its initial
energy. Thus, in case of only
thermally driven wind one could encounter a situation where the gas has
lost most of its thermal energy, the inner shock catches up with the outer shock
and only the momentum of the free wind drives the outflow. This is known
as `momentum driven outflow' \citep[see][for evolution of
such outflows]{2008MNRAS.385..783S}.
However, in the presence of non-zero cosmic ray energy, we will find that there
is always a finite separation between the inner and outer shocks and
hence the outflow never transits to the momentum driven case. Further,
owing to a softer equation of state, cosmic rays lose energy slower than
the thermal gas during adiabatic expansion. Hence
even when radiative cooling of the bubble gas is efficient,
cosmic ray pressure is important and moreover
it dominates the outflow dynamics at the later
stages of its evolution.

Most analytical models of galactic outflows do not consider
the dynamics of the inner shock at $R_1$ \citepalias[but see][]{2008MNRAS.385..783S}.
We follow its dynamics using
simply the jump condition across the inner shock boundary in a two
fluid model assuming a strong shock. The evolution equation for $R_1$ is
given by \citep[][and Appendix~\ref{appendix1}]{1983ApJ...272..765C}
\begin{equation}
P_b +P_c =  \frac{2}{\gamma_s+1} \f{\dot{M}_w(t_e) }{4 \pi R_1^2 v_w}\left[ v_w - \dot{R}_1\right]^2,
\label{R1eqn}
\end{equation}
with 
\begin{equation}
\frac{2}{\gamma_s+1}=\frac{3\left(1+2\frac{P_{c}}{P_b}\right)}{4+7\frac{P_{c}}{P_b}}.
\label{gammas_eqn}
\end{equation}
Here, $v_w$ is the asymptotic speed of the free wind material before
it encounters the inner shock and related to the SNe energy by
$L_0(t)=\dot{M}_w(t) v_w^2/2$, with the mass outflow rate $\dot{M}_w(t)$
is obtained from Eq.~\ref{eqn_eta}.

The mechanical luminosity, $L(t)$, and the cosmic ray luminosity
at the inner shock, $L_c(t)$, fed into the wind bubble
are related to the SNe explosion energy and hence the star formation
rate of the host galaxy.
Our model for calculating this star formation rate is described in
Appendix~\ref{sec_sfr}.
It takes into account the negative feedback due to 
mass loss in the free wind following \citet{2014NewA...30...89S}.
However, we make some modifications
that are needed in order to get a more accurate time resolved star 
formation rate in the low mass galaxies.
The resulting evolution of the SFR for a few 
sample halo masses and collapse redshifts
incorporating SNe
feedback, are shown in Fig.~\ref{fig_sfr}, 
where we can see a significant suppression of the SFR for low mass galaxies.
This model of star formation is similar in philosophy to the `bathtub model' of
\citet{2013MNRAS.435..999D} and \citet{2014MNRAS.444.2071D},
applied however just after the formation of a new
dark matter halo, either formed by mergers or from a collapsing density peak.
Thus only the first burst of star formation lasting several
halo dynamical time scale is considered.
It does not consider the slow and continuous star formation mode of a galaxy
that arises at a later stage of its evolution due to slow accretion
process. Note that such a slow star formation may not result in an outflow
that escapes the dark matter potential. We will explore such effects in
future, but concentrate for now on the major epoch of star formation.

The luminosities $L(t)$ and $L_c(t)$ are related to
the star formation rate $dM_*/dt$ by,
\begin{equation}
L(t) = 10^{51}~{\rm ergs}\times\epsilon_{\rm w}~\nu~\frac{dM_*}{dt} 
\end{equation}
and
\begin{equation}
L_c(t) = 10^{51}~{\rm ergs}\times\epsilon_{\rm cr}~\nu~\frac{dM_*}{dt}.
\end{equation}
Here, we assume that $\nu$ number of SNe are formed per unit 
mass of star formation
and each SNe produces $10^{51}$~ergs of energy on an average. A fraction
of that energy, $\epsilon_w$ is channeled into the hot bubble as thermal energy
and a fraction, $\epsilon_{\rm cr}$, is transformed in to the cosmic ray energy.
In our model where the SNe first drives a free wind, which then flows
into the hot bubble through a strong inner shock, it is this strong shock
which reconverts the kinetic energy of the free wind back into both 
thermal energy and cosmic rays.
Thus, the total energy efficiency in converting SNe energy to energy available
for driving the outflow is $\epsilon_w + \epsilon_{\rm cr}$. For most of our
calculations we assume $\epsilon_w=0.05$ and $\epsilon_{\rm cr}=0.05$ making
total energy efficiency of 0.10. This efficiency is 2-3 times lower than the
efficiency that has been found in numerical simulations (see 
\citet{2002ApJ...571...40M}
where they obtained 20-30\% efficiency) and hence a conservative value.
Further, we assume a Salpeter initial
mass function for the distribution of formed stars in the mass range
$1-100$~M$_\odot$ that results in one SNe per 50~M$_\odot$ of star formation.

The above equations are simultaneously solved numerically
starting from a set of initial conditions as described in
\citetalias{2008MNRAS.385..783S}.
We follow the outflow dynamics till the outflow peculiar velocity 
decreases to the local sound speed in the IGM. At this point we assume that
the shock would disappear and the outflowing material would mix
with the IGM and expand with the Hubble flow.

\section{Magnetic field dynamics}
\label{sec_mag_field}

It is important to explore the magnetisation of the IGM by outflows.
We have already mentioned that the magnetic fields play an important role in
its interaction with cosmic rays and hence probably in the thermal history of
the universe. 
Galactic outflows are potential candidates to magnetise the IGM.
Here, we wish to 
get a conservative estimate of
the strength of the IGM magnetic fields 
resulting from our outflow models. 
For simplicity, we treat evolution of magnetic fields
separately and do not consider any dynamical effects of magnetic fields
on the evolution of the outflow.
Note that, adding the dynamical effect of magnetic fields can further
aid in driving outflows. However, we have adopted a more conservative approach
in present work as 
the magnetic pressure is always significantly subdominant to the CR pressure.

We assume that micro-Gauss level magnetic
fields are present in the ISM of the high redshift protogalaxies when
outflows begin. 
These fields themselves would be generated by turbulent dynamo processes
in the galaxy \citep{2005PhR...417....1B,2008RPPh...71d6901K,RSS88,2015MNRAS.450.3472R}.
The ionised plasma in the wind would
be coupled with this magnetic field and transport the magnetic flux
out of the galaxy with the outflowing material.
In order to calculate the strength of magnetic fields that can be deposited
in the IGM via this process we follow \citet{2006MNRAS.370..319B}
where they considered magnetising the IGM by thermally driven winds.

There are two possible scenarios in the co-evolution of magnetic fields
and outflows. In the first case that we refer as `conservative' model,
the magnetic fields from galaxies are injected into the outflowing
material and just get diluted as they expand with the outflow, obeying
the magnetic flux freezing condition. 
The evolution of magnetic fields in such models is governed by following equation
\citep{2006MNRAS.370..319B},
\begin{equation}
\frac{dE_B}{dt}=\dot{E}_{B_{in}}-\frac{1}{3}\frac{\dot{V}_w}{V_w}E_B.
\end{equation}
Here, $E_B$ is the total magnetic energy in the wind bubble; $V_w$ is the
volume of the wind bubble. Further, $\dot{E}_{B_{in}}$ is the magnetic energy
injection rate in the bubble from the galaxy
and we take
\begin{equation}
\dot{E}_{B_{in}}=\epsilon_{B_{in}}\frac{\dot{M}_w}{\bar{\rho}_{in}}
=\epsilon_{B_{in}} 4\pi R_1^2 v_w
\end{equation}
with the density of the free wind material is given by
$$
\bar{\rho}_{in}=\frac{\dot{M}_w}{4\pi R_1^2 v_w} 
$$
and the magnetic energy density that is injected in the wind bubble
is 
\begin{equation}
\epsilon_{B_{in}}=\frac{B^2}{8\pi} \left(\frac{\bar{\rho}_{in}}{\bar{\rho}_{\rm ISM}}\right)^{4/3}.
	\label{eqn_B}
\end{equation}
$B$ is the magnetic field inside the star forming galaxy.
The value of $B$ ranges from $\sim 10-20\mu$G in nearby spirals
to values of $B\sim 50-100\mu$G in nearby starburst galaxies and in
barred galaxies \citep[see][for a review and references]{2016A&ARv..24....4B}.
There is also tentative evidence from high-$z$ Mg~{\sc ii} system
that $z\sim 1$ galaxies are already magnetized to current levels
\citep{2008Natur.454..302B,2014ApJ...795...63F}.
Thus it seems reasonable to adopt $B\sim 10-20\mu$G for the galaxies
at high redshift which are driving outflows.
Further, $\rho_{\rm ISM}$ is the density of the ISM gas
taken as 1000 times the average baryonic density of the halo.

In the above `conservative' case we ignore the possible amplification
of magnetic fields inside the outflow and thus we get the minimum possible
value of magnetic fields that can be inputted into the IGM via outflows.
In a more optimistic
model~B we take into account possible amplification of magnetic fields in
the wind plasma due to shear flow and turbulence of the wind material.
The characteristic
time scale for this process is roughly given by \citep{2006MNRAS.370..319B},
\begin{equation}
\tau_{\rm{ eff}}^{-1} = f_s\frac{\dot{R}_s - H(t)R_s}{R_s} =
f_s\left(\frac{\dot{R}_s}{R_s} - H(t)\right)
\label{taueff}
\end{equation}
with $H(t)$ is the Hubble parameter. 
Here, $\dot{R}_s - H(t)R_s$ is the peculiar velocity
of the outflow with respect to Hubble flow and 
dividing it by $R_s$ gives a measure of the velocity
shearing rate in the outflowing material. Such a velocity
could reflect itself in turbulence, which amplifies
the field, due to a small-scale turbulent dynamo, 
on the turbulent eddy turn over timescales
\citep{Kaz68,2005PhR...417....1B,2015JPlPh..81e3902B}.
In such a dynamo the smallest eddy which has a magnetic
Reynolds number above a critical value decides the rate
of initial amplification. The fudge factor $f_s > 1$ gives
how much faster such an amplification occurs with respect to the 
shear time-scale. It will depend on the poorly known 
properties of the turbulence in the hot bubble
and could be much greater than unity, if the smallest super critical
eddy is much smaller than the bubble size.
We assume $f_s \sim 1$ as a conservative estimate below. 
Including this amplification, the change in magnetic energy
can be calculated from
\begin{equation}
\frac{dE_B}{dt} = \dot{E}_{B_{in}}+ \left(\frac{1}{\tau_{\rm{eff}}}-\frac{1}{3}\frac{\dot{V}_w}{V_w}\right)E_B
\end{equation}
which 
for $R_s \gg R_1$,
can be simplified to 
\begin{equation}
\frac{dE_B}{dt} = \dot{E}_{B_{in}} - H(t)E_B.
\end{equation}
This would provide us a more optimistic value of the magnetic fields with
which IGM can be seeded via outflows.

\section{Structural characteristics of outflows}
\label {sec_wind_profile}
\begin{figure*}
\centerline{
\epsfig{figure=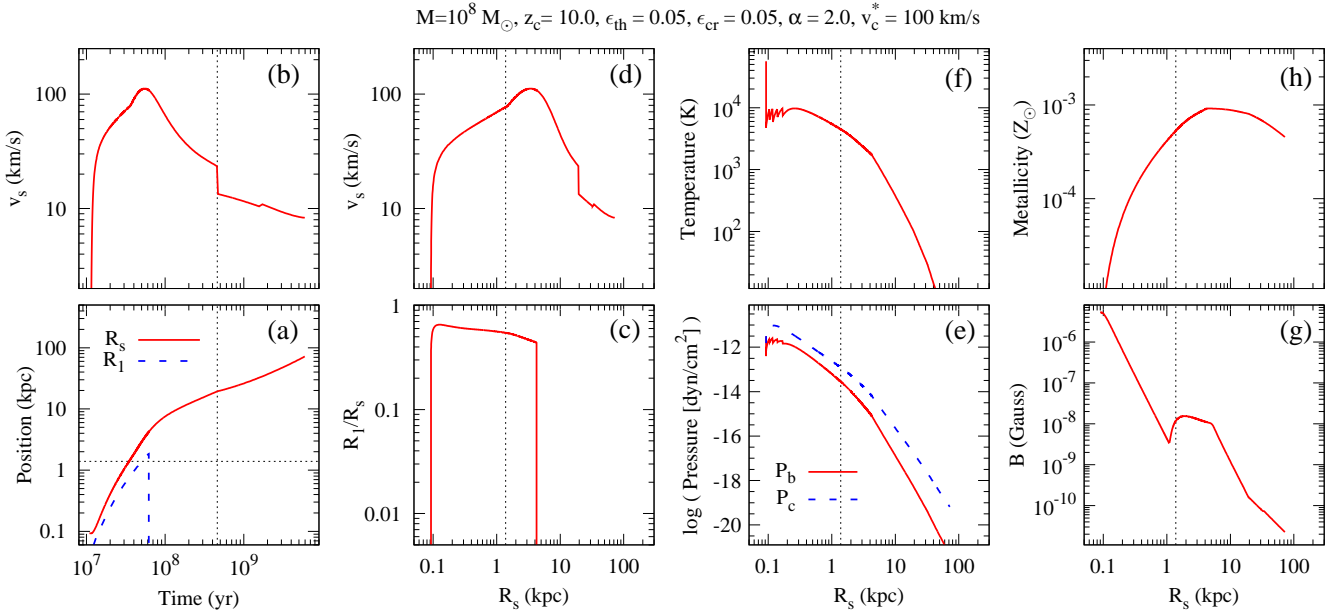,width=8.5cm,angle=-90.0,%
}}
\caption[]{
The characteristics of an outflow originating from a galaxy with total mass $M=10^8 M_\odot$ that has collapsed at $z_c=10$. We show in panel~(a) the position
of inner ($R_1$) and outer shocks ($R_s$) as a function of time.
In panels~(b) and (d) we have
shown the velocity of the outer shock as a function of time and its position ($R_s$) respectively. Panel~(c) shows the ratio between the positions of inner and
outer shocks as a function of $R_s$. The thermal and cosmic ray pressures
as a function of $R_s$ are
plotted in panel~(e) with solid and dashed lines respectively. The evolution
of the temperature and the metallicity of the hot bubble gas are shown in
panels~(f) and (h) respectively. In panel~(g), the evolution of the magnetic
field in the outflowing gas for our conservative model
is plotted as a function of the outer shock position.
The horizontal dotted line in panel~(a) and vertical dotted lines in
panels~(c) to (h) mark the position of virial radius of the halo.
The vertical dotted lines in panel~(a) and (b) show the time when
the outflow merges into Hubble flow.
}
\label{fig_profile}
\end{figure*}

Having set up the machinery for the evolution of outflows from star
forming galaxies, driven by the thermal and non-thermal pressures and the
spreading of magnetic fields and cosmic rays via such outflows, we 
present here some
important characteristics of individual outflows resulting from our
numerical calculations. We first focus on two representative masses of host galaxies
in order to show the general trends in our models. 

\subsection{Outflow from dwarf galaxies}

We begin with the study of outflows from a host
galaxy of total mass $M=10^8$~M$_\odot$ that has collapsed at $z_c=10$.
The star formation history of such galaxies is shown by the solid red curve
in Fig~\ref{fig_sfr}. 
As we can see from this figure, in such small mass galaxies, due to strong 
negative SNe feedback or a large $\eta_w$,  star formation
completely stops after $10^7$~yrs. Hence, it is essential
to investigate whether such 
strongly suppressed and brief duration of star formation can at all
produce enough SNe energy to
drive a galactic scale outflow from the galaxy.
The important characteristics of the outflow originating
from this particular galaxy are shown in Fig.~\ref{fig_profile}. 
Indeed we see in 
Fig.~\ref{fig_profile} that a galactic scale outflow has resulted
from such galaxy. 

\begin{figure*}
\centerline{
\epsfig{figure=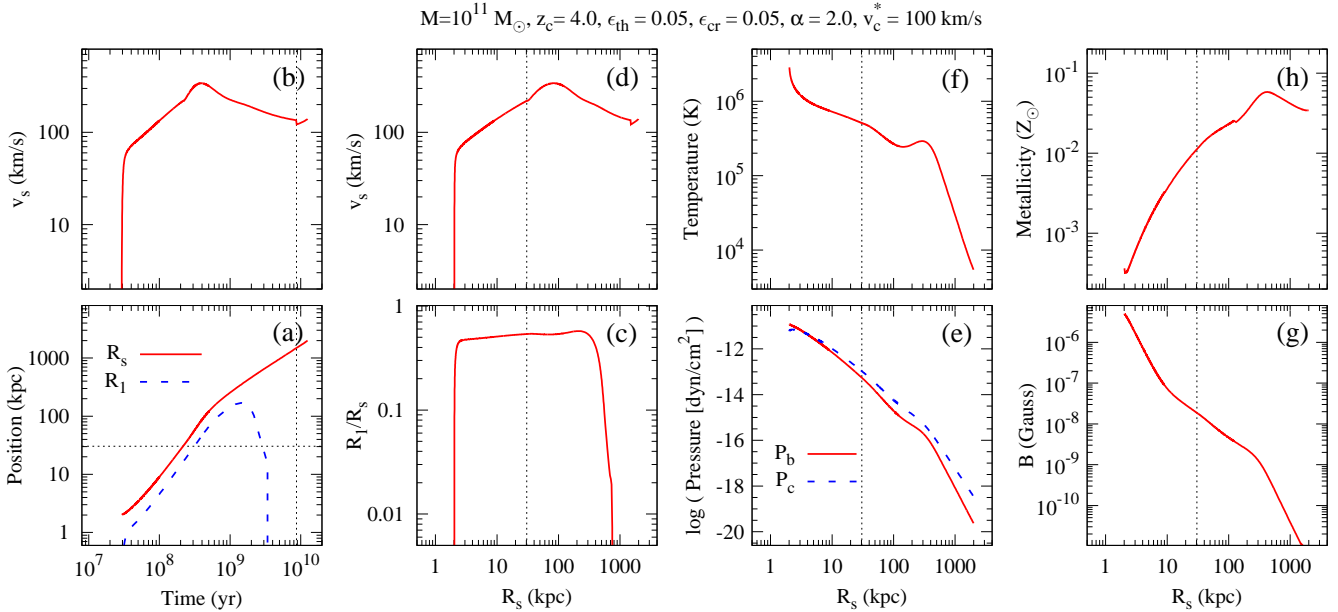,width=8.5cm,angle=-90.0,%
}}
\caption[]{
Same as Fig.~\ref{fig_profile} but for a halo of mass $M=10^{11}$~M$_\odot$ and collapse redshift $z_c=4$.
}
\label{fig_profile11}
\end{figure*}

It is clear from panel~(a) of Fig.~\ref{fig_profile},
where we plot the location of inner and outer shocks as a function
of time,
that the outflow crosses
the virial radius of the halo at $\sim 3\times 10^7$~yrs, goes far beyond
the virial radius (shown by horizontal line) before it mixes with the IGM. The final physical radius of the
outflow is 20~kpc before it is frozen into the Hubble flow after
$\sim 3\times 10^8$~yrs from the collapse of the halo
(as indicated by the vertical lines in panels~(a) and (b) of Fig.~\ref{fig_profile}).
Note that the halo
has a virial radius of 1.4~kpc. Therefore, the outflow originating from the
host galaxy is able to pollute IGM with metals, magnetic fields and CRs 
upto $\sim 15$ times of its virial radius. The outflow
achieves a maximum speed of 110~km/s just after crossing the virial
radius of the halo (panel~(b) and (d) of Fig.~\ref{fig_profile}).
The sudden increase in the velocity after crossing the virial
radius is due to rapid change in the assumed gas density profile outside the halo.

It is interesting to note that even though
star formation ceased after few times $10^7$~yrs
the inner shock still persists 
upto a time $ \sim 5 \times 10^7$~yrs (panels~(a) and (c)). This is due
to the fact that it takes a finite time for 
the effects of energy release by SNe to be felt at the inner shock.
The inner shock starts from the centre of the galaxy (our initial condition), rapidly
goes closer to the outer shock reaching to a maximum of $0.65 R_s$. Thus,
it never comes very close to the thin shell and the outflow never
transits to a momentum driven stage. It can be seen from panel~(f) of 
Fig.~\ref{fig_profile} that the temperature of the bubble is always less
than $10^4$~K. So in the absence of CRs, the thermal pressure alone
would not have sustained the bubble
region leading to a direct momentum impact of 
the free wind to the thin shell. However,
in presence of cosmic ray pressure this never happens. 
At a very initial stage of the flow the thermal pressure
and cosmic ray pressure were comparable to each other. But soon the
gas loses its thermal energy by radiative cooling due to higher
density of the wind material. At this point, the comic ray pressure
becomes an order of magnitude higher than the thermal pressure even if
initially they were of same order (see panel~(e) of Fig.~\ref{fig_profile}).
This extra cosmic ray pressure completely
governs the further evolution of the outflow keeping it always like
an `energy driven flow'.

We now turn to the spreading of metals and magnetic fields by the outflow.
Panel~(g) of Fig.~\ref{fig_profile} shows the 
magnetic field strength in the outflowing gas, as a function of the 
outflow radius. Here, we have used the
`conservative' model for the magnetic fields evolution,
and assumed $B\sim 10 \mu$G in the galaxy. By the time outflow
reaches the virial radius,  the average strength of the magnetic fields
is $\sim 0.02~\mu$G. It reduces farther reaching to a value of
$0.4~$nano Gauss when the outflow mixes with the IGM.
This field strength could be even larger if we take larger $B$
for the galactic field
as would be appropriate for a star bursting galaxy. 
Note that the `optimistic' model would predict an order of magnitude
higher magnetic field strength.
The implications of such fields will be
discussed in Sec~6.2 and 6.3.
The amount of metals carried by the wind material from
the ISM to the bubble increases with time and hence it increases the
metallicity of the bubble gas. We see from panel~(h) that by this
outflow we can pollute IGM with a metallicity of $Z\sim 0.001$~Z$_\odot$. 

In a nutshell, the star formation and resulting SNe explosions in a
$10^8$~M$_\odot$ halo can produce a galactic scale
outflow of `cold' gas, mostly driven by the cosmic ray pressure and can
pollute a spherical region of radius 20~kpc (i.e. $\sim 15$ times
the virial radius) with metals (metallicity
of the polluted region is typically $\sim 0.001$~Z$_\odot$), cosmic rays and magnetic fields with 
of order a nano Gauss strength when it mixes with the IGM.

\subsection{Outflow characteristic of a galactic scale halo}

We now consider the 
outflow from a more massive halo with mass $M=10^{11}$~M$_\odot$
and collapse redshift of $z_c=4$. The outflow properties are shown
in Fig.~\ref{fig_profile11}. We choose such an example as we will show
later that these halos contribute most to the global volume filling of
the wind material today. The basic picture remains the same as its low mass
counterpart. However, there are certain differences. Owing to its higher star
formation rate, the outflow from the halo extends to more than a Mpc from
the centre of the galaxy before freezing into the Hubble flow
(panel~(a)). It
achieves a maximum velocity of 340~km/s while crossing the virial radius
of the halo (panel~(b) and (d)). The outflow remains `energy driven' through
out its lifetime; the inner shock never catches up the outer shock as can be
seen in panels~(a) and (c) of Fig.~\ref{fig_profile11}.

The most striking difference between outflows from high and low mass galaxies
is in the temperature evolution. In the low mass galaxy the outflowing material
is cold. But in the high mass galaxy the density of hot bubble is low due
to smaller $\eta_w$ making
the radiative cooling less efficient. Thus as can be seen in panel~(f)
of Fig.~\ref{fig_profile11}, the temperature of the hot gas is more
than $10^6$~K when the outflow starts. It decreases afterwards, but never
becomes less than $10^4$~K when the outflow is in active stage.
Further,
the thermal and cosmic ray pressures remains comparable to each other
through out the life time of the outflow (panel~(e)). It is interesting
to understand the evolution of both the pressures. Initially, the thermal
pressure is higher than the cosmic ray pressure.
However, after some
time the cosmic ray pressure becomes higher. Such a feature arises
due to different adiabatic indices for thermal gas and relativistic cosmic ray
component causing pressure of the non-thermal component to decrease
more slowly under adiabatic expansion (see Eqs.~\ref{pressure_energy} and
\ref{pressure_energy_cr}).

As can be seen from panel~(h) of Figs.~\ref{fig_profile} \& 
\ref{fig_profile11}, $10^{11}$~M$_\odot$ halo spreads the metals
over large distances and the resulting gas phase metallicity in the
polluted region can be two orders of magnitude higher than that
of a $10^{8}$~M$_\odot$ galaxy.
The large galaxy produce more metals
through SNe and thus eject more via outflow.
On the other hand, they both magnetise the IGM with similar magnetic
field strength but over different volumes. 
Thus outflows from massive galaxies pollute
a fairly large volume of IGM with hot gas having higher metallicity. But
they contribute in a similar fashion in magnetising the IGM. This may
change if we consider a halo mass (or SFR) dependence in $B$ of Eq.~\ref{eqn_B}.

Further, note that in panel~(g) of
Figs.~\ref{fig_profile} and \ref{fig_profile11}, the evolution of the
magnetic fields has been obtained for our `conservative' models.
The more `optimistic' model~B produces an order of magnitude higher magnetic
field in the IGM. We will discuss its implication on the thermal history
of IGM in section~6.2.

\subsection{Comparison with Previous models}
In this section we compare both cosmic ray and thermally driven outflows
with models of \citetalias{2008MNRAS.385..783S} that considered only the thermally driven outflows
and a fixed mass loading factor $\eta_w=0.3$.
As already mentioned, there are several improvements in present models
compared to the model presented in \citetalias{2008MNRAS.385..783S}.
Firstly, the prescription for the star formation now takes into account
the negative feedback arising from the SNe driven winds.
Secondly, we also take into account the increased mass loading factor
$\eta_w = (v_c/v_c^*)^{-2}$ in small galaxies in the outflow
dynamics as adopted in our star formation model.
Finally, the
winds are now driven by both thermal and cosmic ray pressures. As
described before, the
first two effects have opposite influence on the
wind dynamics compared to the last one.
In this section we compare different models
in more detail.

\begin{figure*}
\centerline{
\epsfig{figure=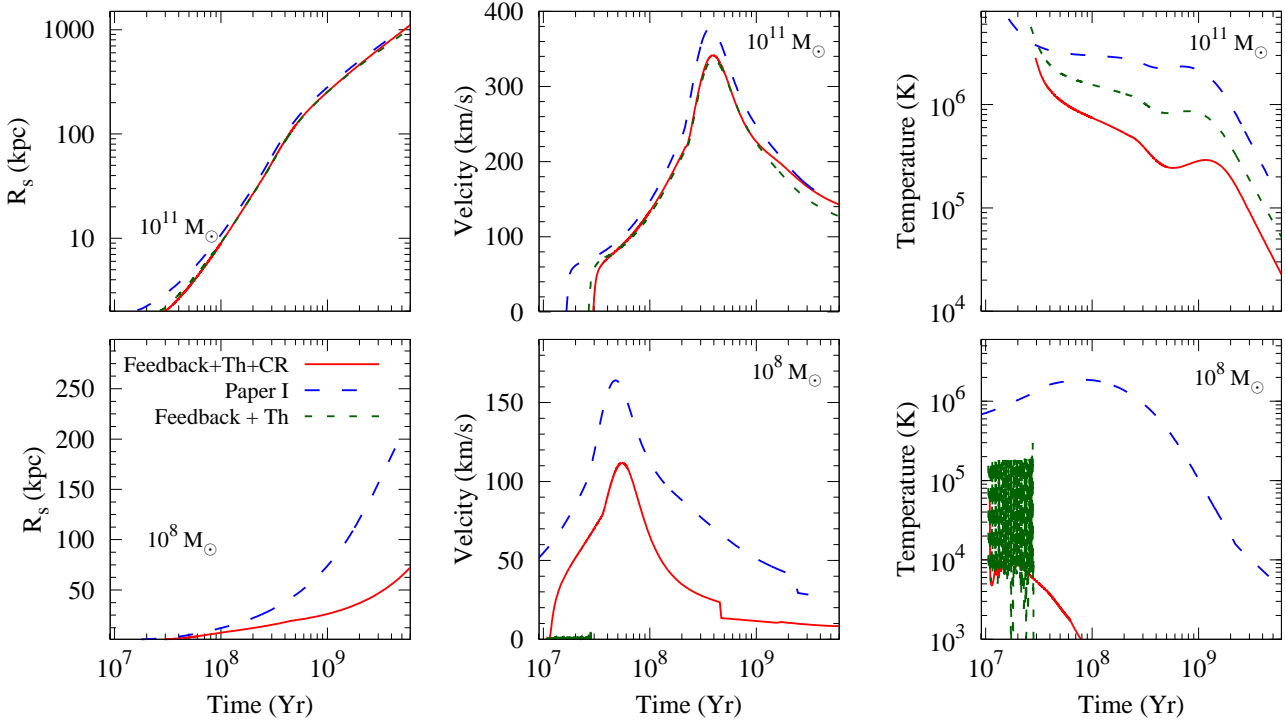,width=10cm,angle=-90.0,%
}}
\caption[]{Comparison of old models of \citetalias{2008MNRAS.385..783S} with new models for galaxies
with masses $10^8$~M$_\odot$ (bottom panels) and $10^{11}$~M$_\odot$ (top panels).
We show the outer shock positions (left panels), shock velocity (middle
panels) and temperature of bubble gas (right panels) as a function
of time. The solid red lines are for present models that consider SNe
feedback in star formation and both cosmic rays and thermal gas drive
outflows. Predictions by the models of \citetalias{2008MNRAS.385..783S} that assumed only thermally
driven outflows and did not take into account SNe feedback in star formation
are shown by blue long dashed lines. Finally, outflows that are generated
by only the hot thermal gas due to star formation with SNe feedback are
shown by dark green short dashed lines.
}
\label{fig_comp}
\end{figure*}

In Fig.~\ref{fig_comp}, we show the properties of thermally driven
outflows with models discussed in \citetalias{2008MNRAS.385..783S} by blue long dashed curves.
The corresponding curves for models
that includes SNe feedback,
the increased mass loading but no effect of cosmic rays,
are given in short dashed dark-green lines.
In both the cases, we assume $\epsilon_w=0.1$
and $\epsilon_{\rm cr}=0$. The effect of the SNe feedback,
increased mass loading, but now including the cosmic rays in the outflow dynamics 
keeping the total efficiency the same (i.e. $\epsilon_w + \epsilon_{\rm cr}
=0.1$), is shown as red solid lines.
The bottom panels consider
outflow from a $10^8$~M$_\odot$ galaxy
while the top panels are for $10^{11}$~M$_\odot$ halo.
It is clear from the figure that the outflow escapes the
low mass galaxy for the constant $\eta_w$ model of \citetalias{2008MNRAS.385..783S}.
However, including supernova feedback
in star formation and increased $\eta_w$ prevent the outflow from
taking off in such galaxies.
This is reflected by the fact that
the green short dashed curves are hardly noticeable in the first two
columns of bottom panels
of Fig.~\ref{fig_comp}.
This is because
of (i) the reduction in the star formation (especially due to the larger
$\eta_w$ in dwarf galaxies) and (ii) increase in the mass
loading that leads to higher density and hence rapid cooling of the hot bubble gas.
It is also evident from the figure that one can have outflows in such
halos when we include the cosmic rays in the outflow dynamics (solid red curves)
keeping the total energy efficiency the same. 
Thus we conclude that the presence of CRs enables outflows even
in small mass galaxies
with a large mass loading.
Further, the outflowing gas will have low temperature due to efficient
radiative cooling (temperature
is $< 10^4$~K as can be seen in the left bottom panel of Fig.~\ref{fig_comp}).

For large galaxies with $M=10^{11}$~M$_\odot$, 
outflows do escape the dark matter potential in all
three cases as described above. However,
the cosmic ray driven outflows with same energy 
efficiency reaches much larger distance (see top panels of Fig.~\ref{fig_comp}). This is because the cosmic ray 
pressure drops slower with radius than the thermal pressure due to a different adiabatic
index. Thus cosmic ray pressure is able to push the outflowing
gas further away compared to the one driven by thermally driven outflows.
This can be seen from the velocity profile as plotted in the top
middle panel of Fig.~\ref{fig_comp}. Further, the temperature
of the bubble gas is also lower by an order of magnitude in case of cosmic
ray driven outflows making it more plausible for detection of metals.

\section{Global influences of outflows}
\label{sec_global}
\begin{figure*}
\centerline{
\psfig{figure=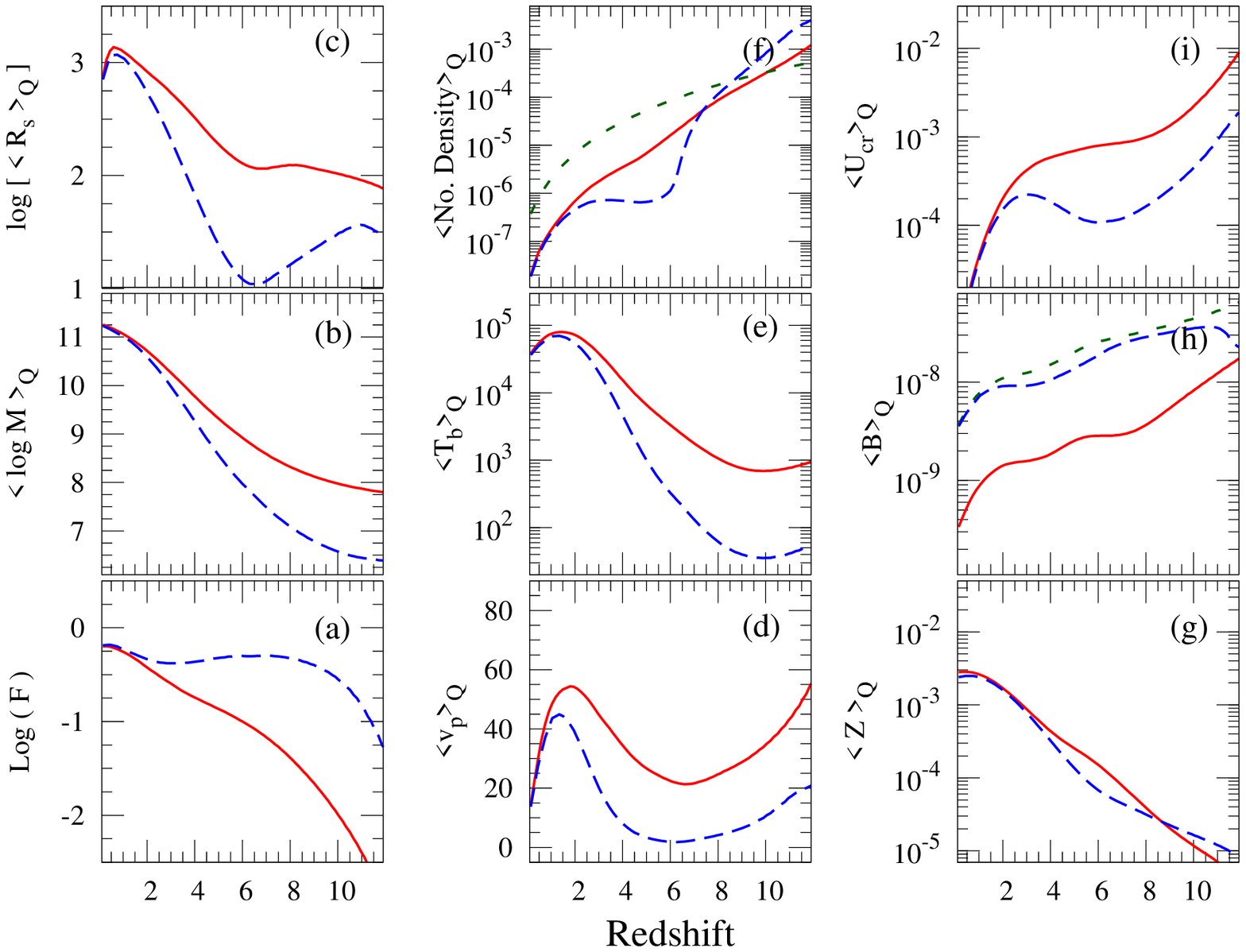,width=12.5cm,angle=-90.0,%
}}
\caption[]{
Global properties of outflows resulted from atomic cooling model (solid lines)
and molecular cooling models (dashed lines). We show
volume filling factor $F$ in panel~(a). The porosity weighted average
halo mass (in unit of M$_\odot$), outflow radius ($R_s$ in kpc),
peculiar velocity ($v_p$ in km/s) and bubble
gas temperature ($T_b$ in K) are shown in panels~(b), (c), (d) and (e) respectively.
In panel~(f) we have compared the average number density per cc of the gas
inside the bubble with
	the IGM number density density at that redshift (dotted curve). The amount of metals (in unit of Z$_\odot$)
 and magnetic fields in conservative models ($B$ in Gauss) that can be put in the IGM are shown
in panel~(g) and (h) respectively. Result for the optimistic models with 
molecular cooled halos is shown by dotted line in panel~(h). Finally in panel~(i) we show the excess
cosmic ray energy density ($U_{cr}$ in eV/cm$^3$).
}
\label{fig_profileg}
\end{figure*}

In previous section we have investigated several key characteristics
of outflows originating from the high redshift star forming galaxies.
Here we show the global impact of outflows on the IGM
at different redshifts. The most interesting physical quantity is the
total volume of the universe/IGM that is affected/polluted by these
outflows. In this respect we define volume filling factor as
$F=1-\exp [-Q(z)]$ with porosity $Q$ is defined as
\bea
Q(z) = \int \limits_{M_{\rm low}}^{\infty} \de M \int \limits_z ^{\infty}
\de z' \f{\de^2 N(M,z,z')}{\de z'\de M} \f{4}{3}\pi \left[R_S(1+z)\right]^3.
\label{eqn_Q}
\eea
Here, ${\de^2 N(M,z,z')}/{\de z'\de M}$ is the comoving number density of dark
matter halos having mass between $M$ and $M + \de M$ that are formed in redshifts range $z_c$ and $z_c + \de z_c$ and surviving till redshift $z$ without
being merged with other halo.  We use the modified Press-Schechter (PS)
formalism of \citet{1994PASJ...46..427S}
to calculate this number density. In particular, it is obtained from
\citep{2000ApJ...534..507C,2002MNRAS.336L..27R,2007MNRAS.377..285S}
\bea
\f{\de^2 N(M,z,z')}{\de z'\de M}~ \de M~ \de z' &=& N_M(z') \left(\f{\delta_c}{D(z') 
\sigma(M)}\right)^2 \f{\dot{D}(z')}{D(z')}\;\nonumber \\
& \times & \f{D(z')}{D(z)} \f{\de z'}{H(z') (1 + z')}~ \de M.
\label{eqnmPS}
\eea
Here, $N_M(z') \de M$ is the Press-Schechter mass function
\citep{1974ApJ...187..425P},
$\delta_c=1.686$ is the critical over density for collapse, $H(z)$ is the
Hubble parameter, $D(z)$ the growth factor for linear perturbations and
$\sigma(M)$ the rms mass fluctuation at a mass scale~$M$. 

Note that Eq.~\ref{eqnmPS} does not take into account the survival of the
hot bubble when two halos merge to form a bigger halo and thus using it in Eq.~\ref{eqn_Q} would provide a lower
limit on $Q$ or $F$. On the other hand if we use simple derivative of any mass
function as  the formation rate of halos and use it as
${\de^2 N(M,z')}/{\de z'\de M}$, we would get an upper limit on the $F$ or $Q$
as we do not consider the merging of outflows.
To explore this we use derivative of Sheth-Tormen mass function
\citep{1999MNRAS.308..119S}
to calculate the formation rate of halos and hence the volume filling
factor and associated properties of outflows.

The lower mass limit ($M_{\rm low}$) in Eq.~(\ref{eqn_Q}) is determined from the
physical conditions required to host star formation. Before reionisation,
gas inside a dark matter halo can cool via atomic hydrogen if virial temperature ($T_{\rm vir}$)
of the halo is greater than $10^4$~K. In the presence of molecular hydrogen,
a halo with virial temperature as low as 300~K can cool and host star formation
\citep{1997ApJ...474....1T,2000ApJ...534...11H}.
In what follows we consider two models: (i) atomic cooled models where we assume
$M_{\rm low}$ corresponds to $T_{\rm vir}=10^4$~K and (ii) molecular cooled models
where we take  $M_{\rm low}$ corresponds to $T_{\rm vir}=300$~K. Further,
in the ionised regions of the universe due to radiative feedback,
a halo can host star formation only
if its circular velocity is more than about 35~km/s due to increase of Jeans mass
\citep[see][for details of radiative feedback]{2002ApJ...575..111B,2008MNRAS.385..783S,2014MNRAS.443.3341J}.
Thus in order to know whether a halo can host star formation or not
we need to follow the reionisation history of the universe self-consistently.
We follow \citet{2007MNRAS.377..285S}
in order to calculate the ionisation state
of the universe. At this point we wish to note that all our self-consistent
reionisation models are compatible with available observations, namely
reionization redshift $z_{re}\gtrsim 6$ and the resulting electron
scattering optical depth is within $\tau_e=0.058\pm 0.012$
\citep{2016A&A...596A.108P}.

Further for any physical quantity related to outflows, we calculate porosity weighted average
as
\bea
\langle X\rangle &=& Q^{-1} \int \limits_{M_{\rm low}}^{\infty} \de M 
\int \limits_z ^{\infty} \de z'~ \f{\de^2 N(M,z,z')}{\de z'~\de M}\times \nonumber \\
& & ~~~~~~~~~~~~~~~~~~~~~~~~~~~\f{4}{3}\pi 
\left[R_S(1+z) \right]^3 X,
\label{eqnavg}
\eea
$X$ being the physical quantity. If one does not include $Q^{-1}$ factor it
would correspond to the mean value of $X$ in the IGM.

\subsection{Atomic cooling models}

We start by showing results for our `atomic cooled' models with
Press-Schechter mass function. In Fig.~\ref{fig_profileg} we show
various physical quantities related to outflows as a function of redshift by solid curves.
In panel~(a) we show the most relevant quantity, the volume filling factor $F$
of the bubbles.
Initially at $z=10$ the filling factor is small $\sim 0.01$, gradually increases to 0.26 at $z=3$
and finally reaches to 0.6 by $z=0$. Thus a minimum 60\% of the universe can be
filled with outflows from atomic cooled halos by today.
The average halo mass contributing to the volume filling is plotted
in panel~(b). Above $z=6$ the average mass of haloes contributing
most in volume filling is $\sim 10^8$~M$_\odot$. During this
period the average bubble radius is 100~kpc (panel~(c)).
At later times both the average mass and bubble size increase
slowly due to two reasons. Firstly, the number of higher mass galaxies
are increasing in a hierarchical structure formation scenario.
Secondly, the existing outflows are getting more time to grow in size.
At $z=0$, massive galaxies with average mass $\sim 10^{11}$~M$_\odot$
having outflows of Mpc scale are responsible for polluting the IGM.

The porosity weighted peculiar velocity at different redshift is plotted
in panel~(d) of Fig.~\ref{fig_profileg}. Initially, most of outflows
are in the early active stage having higher individual outflow velocity; thus
the average peculiar velocity is also as high as 50~km/s. However, the outflows
are coming from small mass galaxies. Hence they have a lower average
temperature of $10^3$~K (see panel~(e)). As the time passes,
these outflows grow older and their speed reduces.
But as new massive galaxies
started collapsing, the peculiar velocity raises to 55~km/s at $z=2$.
The average temperature of the bubble also shows a peak of $10^5$~K at
the same time. The average density of the bubble is shown in panel~(f)
by the solid curve and most of the time it stays bellow the average
IGM density (shown by the dashed line) at that redshift. This is expected
as the hot bubble sweeps up the IGM into a shell
creating a low density, high temperature regions.

In passing we note that several porosity weighted average physical
quantities do not show monotonic behaviour with redshift such
as porosity weighted peculiar velocity and radius of outflows.
A number of counter veiling effects are responsible for such non-monotonic
behaviour of the porosity weighted peculiar velocity ($v_p$). 
Although the characteristic mass of collapsing halos increases monotonically
with decreasing redshift, the corresponding stellar to halo mass ratio is
not monotonic with mass because of the reionization, supernovae and 
AGN feedbacks. Also as outflows expand, their velocity decreases, which
is reflected by corresponding decrease in porosity weighted $v_p$ 
for $z > 6$. Then after reionization, small mass halos lose their importance, 
more and more massive galaxies with younger outflows start to dominate 
and increase the average peculiar velocity till $z \sim 2$. After this star 
formation from very massive galaxies is being suppressed due to AGN 
feedback, while outflows from normal galaxies are becoming older, and
the weighted $v_p$ decreases again.

One of the reasons for studying outflows is to understand how efficiently
they can pollute IGM with heavier elements, magnetic fields and CRs. The
metallicity and magnetic field evolution of the IGM are shown
in the panels~(g) and (h) of Fig.~\ref{fig_profileg}, respectively. 
As expected the IGM metallicity level increases with time as
more and more SNe explode and resulting metals are put into the IGM via
outflows. Finally, the metallicity of the IGM that can be reached by today
is $\sim 10^{-2}$~Z$_\odot$ (panel~(g)). Note that these metals are distributed
in 60\% volume of the universe. In the same volume, the average magnetic field strength
that can be seeded by outflows is 
$0.4 - 4$~nano Gauss at $z=0-3$, for an assumed $B\sim 20\mu$G in galaxies, 
for our conservative model. Models with optimistic magnetic fields evolution 
predict, $\langle B\rangle \sim 2 -20$ nano Gauss, or 
an order of magnitude higher magnetic fields seeded by this mechanism,
as can be seen from the short dashed line of panel~(h) in Fig.~\ref{fig_profileg}.

Finally, it is interesting to investigate the magnitude of the
porosity weighted cosmic ray energy density
that is still remaining in the plasma. We show this in panel~(i)
of Fig.~\ref{fig_profileg}. Before, $z= 6$ the remaining CR energy density is
as greater than $10^{-3}$~eV~cm$^{-3}$. However, only 10\% of the universe
is filled with these cosmic rays. Even at low redshift the energy density
of cosmic rays does not reduce
much. At $z=3~(1)$ the excess energy density of in the cosmic rays is
$5\times 10^{-4}$~eV~cm$^{-3}$ ($10^{-4}$~eV~cm$^{-3}$) in 26\% (55\%)
volume of the universe. We will discuss
this further in the following section and compare this energy density with the thermal energy density
of the IGM.

We note that there is not much qualitative change in the globally 
averaged physical properties associated between our current model, where 
outflows are driven by both thermal and cosmic ray pressure,
compared the models of \citetalias{2008MNRAS.385..783S},
where we considered outflows to be only thermally driven.
The crucial difference however, is that the current models have taken
into account  also the negative feedback on star formation due to SNe driven
mass loss.
One property which does change is the overall volume filling factor,
which is smaller in the current models, compared to \citetalias{2008MNRAS.385..783S},
because of the overall reduction in star formation due to SNe feedback.
Further, the average temperature
of the outflows has reduced dramatically,
due to enhanced mass loading in the hot bubble,
which will help in the detectability
of the heavier elements in the quasar absorption spectrum.

\subsection{Molecular cooled models}

We now turn to the `molecular cooled' models. As discussed in
\citetalias{2008MNRAS.385..783S} 
outflows from atomic cooled halos can potentially
disturb Lyman-$\alpha$ forest due to their large peculiar velocity and higher
temperature and we showed that molecular cooled halos provide a respite
from this by filling the IGM at very early epoch with smaller outflows.
Note that, in such small mass galaxies the mass loading factor
$\eta_w$ is so high that the star formation ceases with the onset
of supernova explosion, after $t_{\rm SNe} \sim 3\times 10^7$~yr.
Within that period we find that in such galaxies the fraction of
baryonic mass turned into stars (i.e. $M_*/M_b$)  is of order 0.05
at $z=10$. Thus only about 5\% of the baryons 
is converted to stars in such molecular cooled galaxies at high-$z$.
Nevertheless, this does have a significant effect on the volume filling 
at high redshift.
In Fig.~\ref{fig_profileg} we show the global properties of outflows
for such models with long dashed lines. As expected, in these models outflows
volume filled quite early; at $z=8$, more than 75\% of the IGM is filled
by the wind material and hence with metals, cosmic rays and magnetic fields.
The contribution comes mostly from galaxies of masses $10^7$~M$_\odot$
or less with average bubble size less than 30~kpc (see panels~(b) and
(c)). 
Upto $z=3$ the average peculiar velocity of the outflowing material
is less than 20~km/s with average outflowing gas temperature less than
$10^4$~K. Thus, cold gas from the small mass molecular cooled halos
dominates the volume filling of the IGM and is expected not to disturb
the Lyman-$\alpha$ forest. 
Note our results for molecular cooled halos will depend very much on the
assumed value of $M_*/M_b$. This quantity is poorly constrained for such
halos and some models suggest that it could be much less than what we
assume here \citep{2017MNRAS.469.1456V}. Understanding star formation
efficiency in such low mass galaxies at early epoch is essential to accurately
model their effect on global properties.

The other physical quantities such as gas density of bubbles, amount of metals
and magnetic fields put in the IGM
remain similar (panels~(f), (g) and (h)
of Fig.~\ref{fig_profileg} respectively) in both molecular cooled
and atomic cooled models. Note that in Fig.~\ref{fig_profileg}
we have shown the possible magnetisation of the IGM with our
optimistic models as well and that resulted one order of magnitude
higher value compared to conservative model. The excess porosity
weighted cosmic ray energy
density remains one order of magnitude smaller in case of molecular
cooled models for $z\gtrsim 6$. After that the difference
between atomic and molecular cooled models decreases
and finally by $z \lesssim 2$ they become equal.

\subsection{Effect of mass function and formation rates}
\begin{figure}
\centerline{
\epsfig{figure=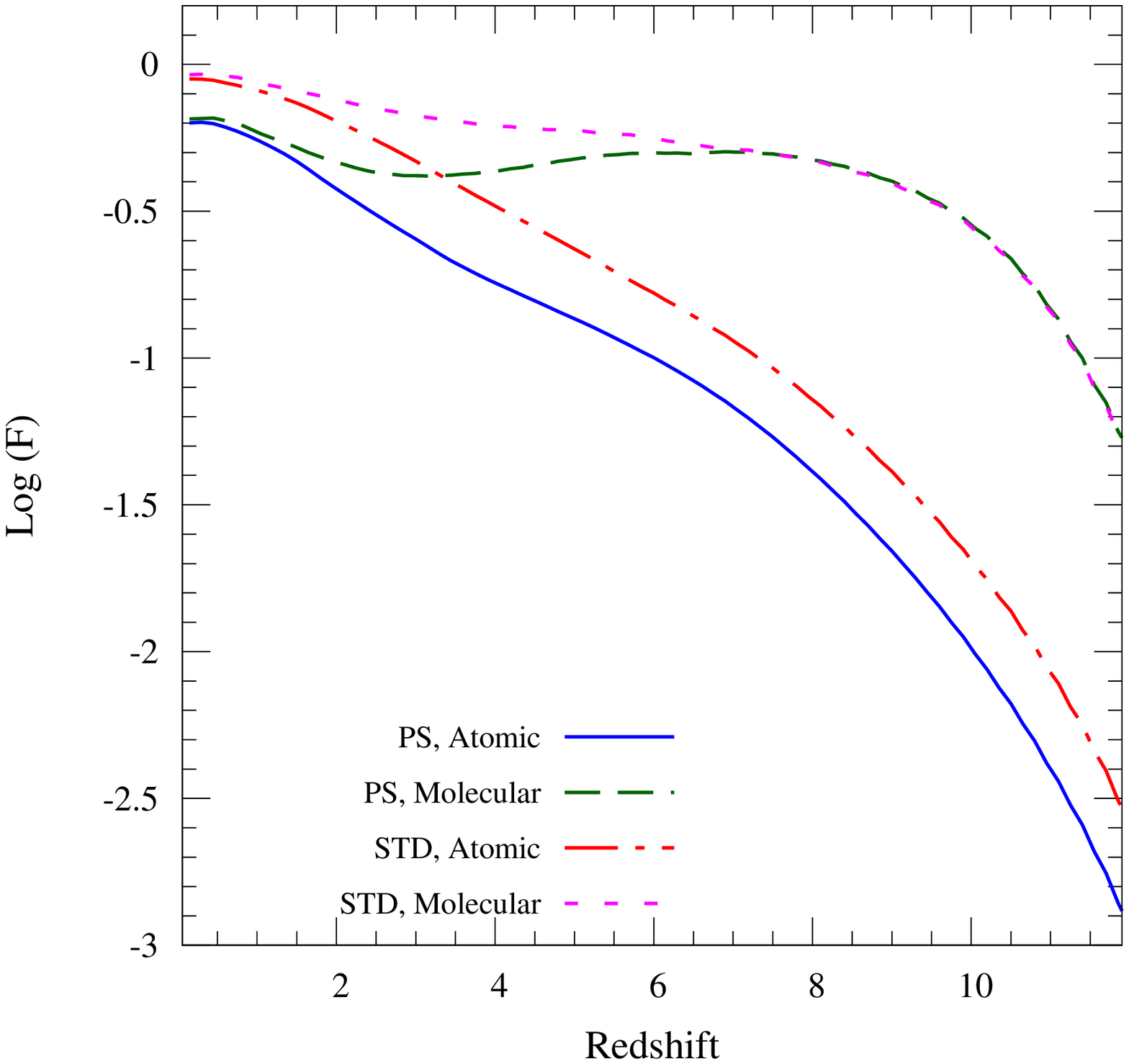,width=7.0cm,angle=-90.0,%
}}
\centerline{
\epsfig{figure=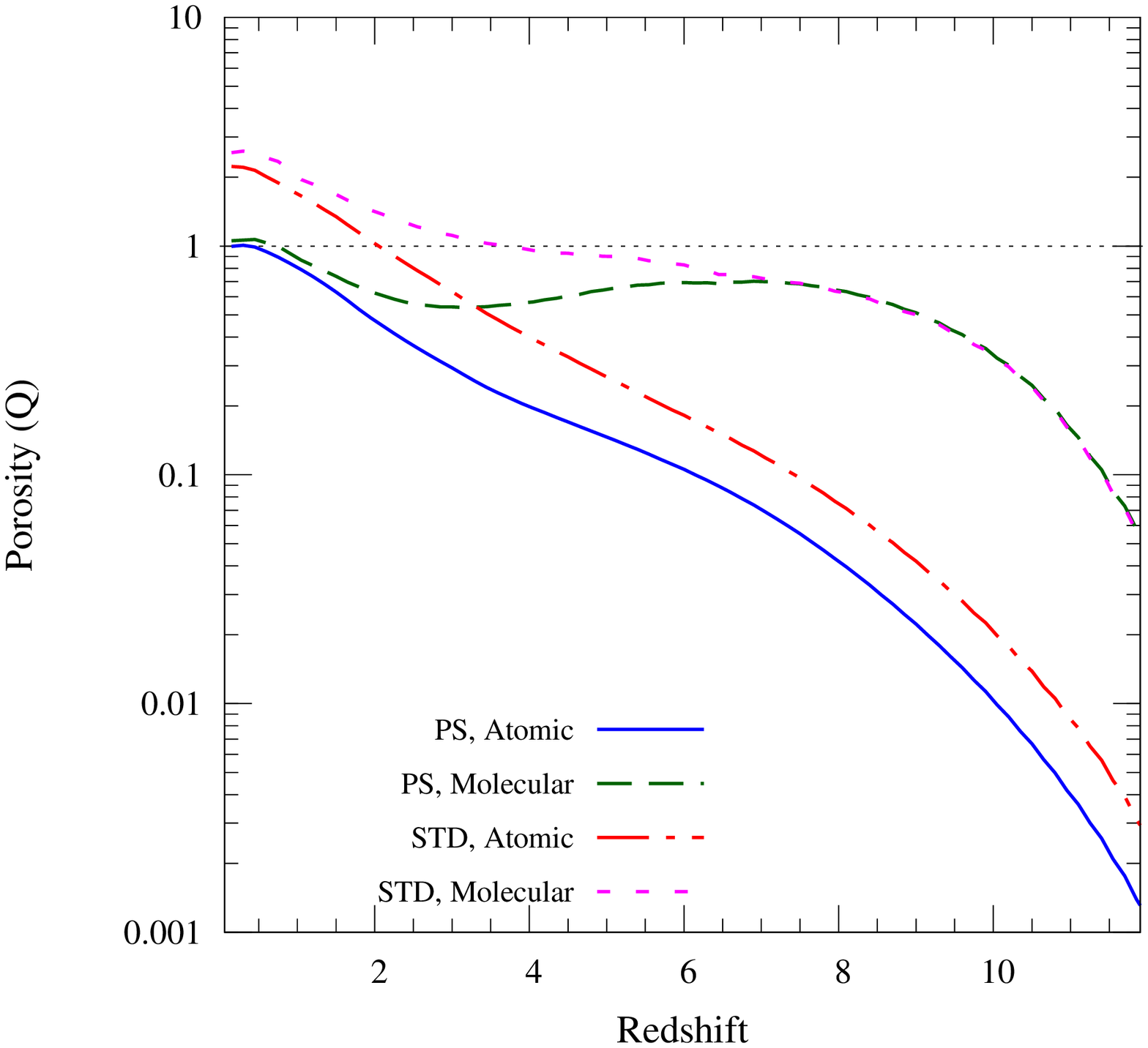,width=7.0cm,angle=-90.0,%
}}
\caption[]{Comparison of volume filling factor (top panel) and porosity (bottom panel) as expected from different mass
functions. Solid blue line and red dotted-dashed line show the filling factor
predicted by atomic cooled models with PS and ST mass functions respectively.
For molecular cooled models the comparison is shown by green dashed (PS)
and magenta dotted (ST) lines.
}
\label{fig_PS_STD}
\end{figure}

Here, we show effects of different halo mass functions on
global properties of outflows.
In Fig.~\ref{fig_PS_STD} we compare the porosity
and the volume filling factor as
obtained from using Sheth-Tormen (ST) mass function with that of
Press-Schechter (PS)
mass function. From the figure it is clear that the Sheth-Tormen mass
function predicts a much larger volume filling of the universe.
For example in atomic cooled models at $z=3$, the ST mass function predicts
60\% volume of the universe is filled with outflows (red dotted-dashed
curve) compared to 35\% with PS mass function (solid blue curve).
Similarly for molecular cooled models at $z=3$, the volume filling
factors are $F=0.65$ and 0.41 for ST and PS mass function respectively.
Note that other physical properties do not change much when using
different prescription for the halo formation rate.

It is interesting to note that  when we 
use the Sasaki formalism to calculate the formation rate of halos 
and include molecular cooled halos, the porosity decreases at $z\sim 6$
before increasing again at $z\sim 2$. In the Sasaki formalism,
one has a survival probability for a halo collapsing at redshift $z_c$
to survive till a later redshift $z$ \citep{1994PASJ...46..427S}.
However when a halo is 
destroyed by merging to form a bigger halo, its outflow will survive, 
if it has gone far beyond the virial radius. Such outflows are not easy to
take into account in a semi-analytical model such as ours, and we have
neglected their contribution to the porosity.
This can lead to 
a decrease in the volume filling factor.
This effect is important when one calculates the volume filling factor
of outflows from
molecular cooled halos which collapse at high-$z$, before reionization. 
On the other hand, models considering derivative of any mass function
would over estimate the volume filling factor by considering
outflows separately from halos after and before merging
\citep{2009NewA...14..591S}.
Thus calculations which employ the formation rate 
using the Press-Schechter-Sasaki formalism and Sheth-Tormen-derivative 
bracket the expected behavior of volume filing factor, 
which is why we have considered both of them in the present work.
Further, in \citet{2009NewA...14..591S} we have tested the sensitivity of the porosity 
not only for various mass functions but also to different prescriptions
for the formation rate of halos. It was shown that the filling factor
as well as other physical quantities were
reasonably insensitive to changes in the mass function.

Thus we conclude that outflows can significantly volume fill the
universe with metals, magnetic fields and cosmic rays.
In passing we note that \citet{2006MNRAS.370..319B} also got similar magnetic field
strength of $10^{-8} - 10^{-12}$~Gauss in the IGM. However, their outflows
never volume filled the universe as their simulation did not include
small mass galaxies.

\section{Implications of outflow models}
\label{sec_impl}
The outflow models discussed here has a number of interesting
consequences, which we discuss below. 

\subsection{Detectability of metals}
In \citetalias{2008MNRAS.385..783S}, we discussed the detectability of
low density high temperature outflowing material in the presence of
meta-galactic UV background through O~{\sc vi} and C~{\sc iv} absorption
lines \citepalias[see section~7 of][]{2008MNRAS.385..783S}.
In particular, we focused on absorption signatures that could arise from
(i) the free wind, (ii) hot bubble and (iii) gas in the shell. The main
difference in our current work is that, in the case of low mass halos
the mass loading is higher and in the case of high mass halos the bubble
temperature is lower when we include the CRs. Moreover, we are also
considering magnetised outflowing material in the current work.
Here we discuss the possible consequences without going into detailed
radiative transport modelling.

In \citetalias{2008MNRAS.385..783S}, it was suggested that the low ion
absorption seen through Mg~{\sc ii}, Na~{\sc i} and Ca~{\sc ii}
transitions can not be produced in the free winds and
could originate from multiphase structure likely
to form inside the outflow due to (i) cold gas ejected from the galaxy along
with the free wind, (ii) dense cloud formed from the Rayleigh-Taylor instability
of the swept up shell or (iii) cold gas clumps formed due to cooling instabilities etc.

\begin{figure}
\centerline{
\epsfig{figure=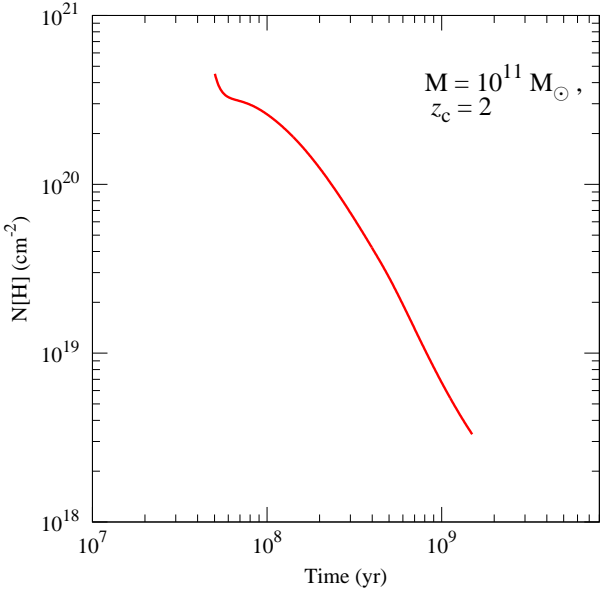,width=8.0cm,angle=-90.0,%
}}
\caption[]{
Hydrogen column density in the free wind for a halo of mass $10^{11}$~M$_\odot$.
It is shown till the star formation lasts.
}
\label{fig_column}
\end{figure}

We first ask, can the free wind driven by CR discussed here be detected
in Mg~{\sc ii} absorption. In Fig.\ref{fig_column} we show the total
integrated hydrogen column density produced by the free wind as a
function of time for a halo of mass $10^{11}$~M$_\odot$. The integrated
column density of total hydrogen inferred is very similar to what we had
for models discussed in \citetalias{2008MNRAS.385..783S}. All the
discussions presented there is applicable here as well. In particular,
to produce low ion absorption lines, we need
multiphase structure with cold clumps originating from the above mentioned
possibilities in addition to the flow structure discussed in this work.

Next we consider the survivability of such cold clouds in the outflow.
As discussed in \citetalias{2008MNRAS.385..783S}, the evaporation time-scale
goes as $T^{-5/2}$. As can be seen in Fig.~\ref{fig_comp}, the bubble
temperature in our present models are typically factor 4 times smaller
than for models considered in \citetalias{2008MNRAS.385..783S}.
This alone makes the evaporation time-scale $\sim 32$ times longer than what
we inferred for the pure thermally driven model without any feedback.
Even when we consider the SNe feedback inclusion of CRs increases the
evaporation time-scale by a factor 5. This suggests that survival of cold
dense clumps in the outflow can be relatively easier in our models with
CR driven flows.
Further, in a magnetised outflow, magnetic fields can drape themselves
around a gas cloud, and reduce evaporation of gas across the field
even further.

\subsection{Thermal history of the IGM}

\begin{figure}
\centerline{
\epsfig{figure=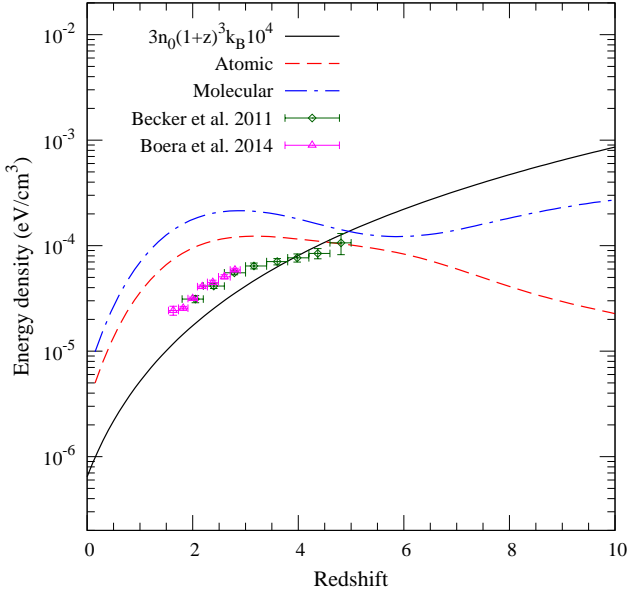,width=8.0cm,angle=-90.0,%
}}
\caption[]{
The thermal energy density of the ionised IGM at mean IGM density with
a temperature $10^4$~K is shown by the solid line. In comparison,
we show the porosity weighted energy density
of the cosmic rays as expected from our atomic cooled models (dashed
line) and molecular cooled models (dotted-dashed line).
 Further, the data points with error bar are the energy
density of the IGM assuming the temperature of the IGM as measured
by \citet{2011MNRAS.410.1096B} (open diamonds) and
\citet{2014MNRAS.441.1916B} (filled triangles).
}
\label{fig_cr_heating}
\end{figure}

We now explore the possible implications of the excess cosmic ray energy
from outflows to the thermal
history of the IGM. As noted earlier, observations of absorption
lines in high redshift quasar spectra reveal that the temperature
of the IGM at $z=2-3$ is few times $10^4$~K
\citep{2011MNRAS.410.1096B}.
This temperature is measured in 
the mildly overdense regions of the IGM with overdensity less than 10 as probed
by the Lyman-$\alpha$ forest.
It is believed that the IGM
temperature keeps some memory of the reionisation process at high-$z$. At the
end of the hydrogen reionisation the temperature of the IGM is set to
$1-3\times 10^4$~K \citep{2000MNRAS.318..817S,2000ApJ...534...41R}
due to photoionisation heating. In absence of any
other heating mechanism after reionisation,
the IGM temperature should drop adiabatically with the expansion of the
universe as $(1+z)^2$. Thus in models where reionization of the universe
occurred earlier than $z=6$ (as expected from observations of quasar spectra)
one would not be able to explain the observed
IGM temperature at $z=2-3$. We need an additional heating mechanism to explain
this temperature floor. Several authors have suggested the late He{\small II} to
He{\small III} reionisation by the harder quasar spectrum
as the reason for this excess temperature
\citep{2008ApJ...682...14F,2009MNRAS.395..736B}.
Here we discuss if excess CR energy from outflows can heat the IGM at $z=2$ to 3.

Note that wherever galactic outflows percolate, they carry
magnetic fields and cosmic rays in addition to metals. Thus, say at $z=3$
for models assuming PS mass function, 50\% of the universe
is filled by the outflows from molecular cooled halos (35\% for atomic cooled
model) with $\sim 10$ nano~Gauss magnetic fields (optimistic model) and excess
average cosmic ray energy density of
$10^{-4}$~eV/cm$^{3}$. With ST mass function for atomic and molecular
cooled models, respectively 65\% and 41\% of the universe
is filled with outflows and similar amount of magnetic fields and excess
average cosmic ray energy density. Note that in standard model of structure formation,
the outflow affected regions are expected
to collapse to form
filamentary structures at latter stages that are likely
to be detected as Lyman-$\alpha$ forests
in the quasar spectrum.  The cosmic ray particles (protons) gyrate
along the magnetic fields lines and generate Alfv\'en waves by a streaming
instability \citep{1969ApJ...156..445K,kulsrud_book}.
These waves damp out and transfer energy to the thermal gas.
The rate of energy transfer to the gas via the Alfv\'en waves is given
by $|v_A.\nabla P_c|$ where $v_A = B/\sqrt{4\pi \rho}$ is the Alf\'ven
velocity ($\rho$ is the density of plasma) \citep{1971ApJ...163..503W}.
The modulus is taken as in this process CRs always lose energy. 
Note that for ionised gas with temperature $T$ the thermal energy density is
given by $E_{th}=3n_Hk_B T$, $n_H$ being the number density of hydrogen
and $k_B$ is the Boltzmann
constant. Thus, time taken to deposit cosmic ray energy of the order of thermal
energy is given by,
$$
t_{cr}= \dfrac {3nk_BT}{|v_A.\nabla P_c|}.
$$
Comparing this with the Hubble time $t_H$ at a given epoch we get,
$$
\dfrac{t_{cr}}{t_H}= H(z)\dfrac {3nk_BT}{|v_A.\nabla P_c|}.
$$
Putting in numbers we find,
\begin{eqnarray}
\dfrac{t_{cr}}{t_H} \approx 0.3 \left(\dfrac{h}{0.7} \right)^{4}
\left(\dfrac{\Omega_m}{0.3} \right)^{1/2}\left(\dfrac{1+z}{4}\right)^6
	\left(\dfrac{T_4}{B_{10 {\rm nG}}}\right) & \nonumber \\
 \times \left(\dfrac{3\times 10^{-4} {\rm eV/cm}^3}{U_{cr}} \right)
  \left(\dfrac{L}{0.1~{\rm Mpc}} \right). & 
	\label{eq-cr-heating}
\end{eqnarray}
Here, $T_4$ is the temperature in the unit of $10^4$~K and $B_{10{\rm nG}}$
is the magnetic field in unit of 10~nano~Gauss. Further we have normalised
this ratio assuming a typical pressure gradient scale
of a filament is $L \sim 0.1$~Mpc \citep{2001ApJ...559..507S} and cosmic ray energy density $U_{cr} = 3\times 10^{-4}$~eV/cm$^3$.
Hence, if the IGM at $z=3$,
has been polluted by outflows with $B$ of 10~nano~Gauss and 
$U_{cr} = 3\times 10^{-4}$~eV/cm$^3$, within 1/3rd of the Hubble
time, the CRs would dissipate
energy to the thermal gas via Alfv\'en waves,
 increasing IGM temperature to $10^4$~K. For even lower redshift, the
process is more efficient.

In Fig.~\ref{fig_cr_heating} we have plotted the volume average cosmic ray energy density
available from our outflow models and compared it with the thermal energy
of the ionised IGM of temperature $10^4$~K. We have also shown
thermal energy of the Lyman-$\alpha$ forests with temperatures as measured
by \citet{2011MNRAS.410.1096B} and \citet{2014MNRAS.441.1916B}.
It is clear that the average cosmic ray energy density around $z\sim 3$
($U_{cr} \sim 10^{-4}$~eV/cm$^3$)
is much higher than the thermal energy density
of the IGM with temperature $10^4$~K, especially for molecular
cooled models. The porosity weighted CR energy density would be
even higher. In the mildly overdense regions of the IGM which produce
Lyman-$\alpha$ forest $U_{cr}$ can be enhanced further by adiabatic compression
to few times $10^{-4}$~eV/cm$^3$.
Magnetic field will also be enhanced due to flux freezing to few tens of nano~Gauss.
Thus one can see from Eq.~\ref{eq-cr-heating}, that
CRs can transfer sufficient energy to the
thermal gas within the Hubble time and potentially heat the IGM around $z=2-4$
to a temperature of $\sim 10^4$~K to erase the
thermal signature of reionization. This conclusion would be
more appropriate for molecular cooled models where outflows
volume filled the universe early in time.

Note that here we have considered
CRs that are accelerated in the inner shocks only. However, we have not considered
the CRs that can be generated in the terminal shock of the outflows as
well as in the structure formation shocks. Simulations predict
cosmic ray energy density of $\sim 10^{-5}$~eV/cm$^3$ that can arise
from structure formation shocks with Mach number of few tens 
\citep{2008A&A...481...33J,2017MNRAS.465.4500P}.
Adding these sources of
cosmic rays would lead to more efficient heating of the IGM by
cosmic rays.

\subsection{IGM magnetic fields}

As noted earlier, $\gamma$-ray observation of nearby blazars (with
$z \sim 0.1-0.2$), have tentatively indicated 
a lower limit for magnetic fields in voids,  
$B_{V}\sim 10^{-16}$~Gauss provided their coherence
length $l_B$ greater than about a Mpc \citep{2010Sci...328...73N,2011MNRAS.414.3566T}. 
For fields with much smaller coherence scale, 
the lower limit increases as $l_B^{-1/2}$ with decreasing $l_B$.
There fields in voids could be primordial arising from processes in the
early universe \citep[][for review]{2013A&ARv..21...62D,2016RPPh...79g6901S}.

However, by such low redshifts, all our models predict a porosity of wind
material greater than unity or volume filling factor greater than
0.6 for randomly distributed sources. 
This implies that outflows could have percolated
even in the void regions by the present epoch. 
Further the porosity averaged value of the magnetic fields in our
models are $\sim$ nano Gauss, and the radius of the outflowing
bubbles is of order Mpc.
The coherence scale of the fields could be smaller than this radius
due to turbulent tangling. And the field strength could be diluted
further in outflows frozen into the expanding IGM in voids. 
Nevertheless, the field strength in our models
are sufficiently large compared to $B_V$, that it is
highly likely that they satisfy the constraints set by the 
$\gamma$-ray observations.
Thus magnetic fields inferred in the voids or in general IGM need not
be of primordial origin. Rather they could be of astrophysical origin,
generated/amplified in the star forming galaxies and transported
from there to the IGM by the outflows created from the supernova
explosions.
One caveat is that large-scale clustering could result in a smaller
volume filling of outflows in the void regions.  It would be of interest to
therefore consider our outflow models in more detail,
in biased proto-void regions.

\section{Discussion and Conclusions}
\label{sec_dc}

In the present work, we have explored how cosmic rays could
aid in driving relatively cold outflows from high redshift galaxies,
and resulting global consequences. The primary sources of energy
for outflows are SNe in the high-$z$ galaxies, and a fraction of this
energy goes into cosmic rays. Outflows also cause a negative
feedback on the star formation rate in a galaxy which can be especially
large in low mass galaxies.
We show that
even with such suppression, low mass galaxies can still drive
an outflow in the presence of cosmic rays.  

Cosmic rays
are accelerated in the SNe shocks and produce cold free winds upto the
inner shock and are then re-accelerated there. These cosmic rays
add extra pressure to the thermal gas that can even drive a cold outflow.
We find that outflows extend to tens of kpc in dwarf galaxies with halo mass $\sim10^8$~M$_\odot$ before merging
with cosmological expansion,
whereas massive
galaxies with halo of masses $10^{11}$~M$_\odot$ have Mpc scale outflows.
Cosmic ray driven outflows can go to a larger radii compare to a purely
thermally driven outflow with same total energy input.
This is because the cosmic ray pressure decreases less rapidly compared to
the thermal pressure with adiabatic expansion (due to its softer equation
of state) and comes to dominate outflow dynamics at large radii.

We showed that these outflows could significantly volume fill the universe with
metals, magnetic field and excess cosmic rays. In particular,
at $z=3$ for the ST mass function with molecular cooled halos, 
65\% of the universe is filled with the outflowing gas
along with metals, magnetic fields of order $10^{-8}$~G and excess
cosmic ray energy density of $\sim 10^{-4}$~eV/cm$^3$. 
All our models predict a volume filling factor of order
unity by $z\sim 0$.

Further, we explore three implications of such outflows. We argued
that due to lower temperature and higher density of the free wind
as well as the bubble gas, the metals spread by the outflow could 
be more easily detected
as absorption line system. Low ionisation metals such as Na~{\sc i} or
Mg~{\sc ii} are likely to be detected in gas clouds embedded
the free wind, whereas others
like C~{\sc iv} or O~{\sc vi} would come from the bubble or from pre-enriched
IGM. Moreover, we showed that the excess energy density in cosmic
rays from these outflows could be transferred efficiently enough
to the thermal IGM gas in presence of magnetic fields.
This can potentially erase
the thermal history of reionisation and heat the IGM to a temperature
of order $10^4$~K as inferred from quasar spectra.
Finally, we find that void magnetic fields of order $10^{-16}$~G
as inferred from blazar observation, need not be of primordial origin,
but could be of astrophysical origin generated in star forming galaxies
and transported by the cosmic ray driven winds to the voids.

Our models which estimate the global impact of outflows on the
IGM, at present do not take account of spatial clustering
of galaxies. In order to have
more realistic scenario, it would be of interests to put the cosmic ray
driven outflows in numerical simulation of structure formation.

\section*{acknowledgements}
We thank the referee for useful comments.
SS thanks UGC for Faculty start up grants. SS also thanks to IUCAA
for its support through associateship programme. 

\bibliography{CR_Wind_Revised}

\begin{thebibliography}{}
\makeatletter
\relax
\def\mn@urlcharsother{\let\do\@makeother \do\$\do\&\do\#\do\^\do\_\do\%\do\~}
\def\mn@doi{\begingroup\mn@urlcharsother \@ifnextchar [ {\mn@doi@}
  {\mn@doi@[]}}
\def\mn@doi@[#1]#2{\def\@tempa{#1}\ifx\@tempa\@empty \href
  {http://dx.doi.org/#2} {doi:#2}\else \href {http://dx.doi.org/#2} {#1}\fi
  \endgroup}
\def\mn@eprint#1#2{\mn@eprint@#1:#2::\@nil}
\def\mn@eprint@arXiv#1{\href {http://arxiv.org/abs/#1} {{\tt arXiv:#1}}}
\def\mn@eprint@dblp#1{\href {http://dblp.uni-trier.de/rec/bibtex/#1.xml}
  {dblp:#1}}
\def\mn@eprint@#1:#2:#3:#4\@nil{\def\@tempa {#1}\def\@tempb {#2}\def\@tempc
  {#3}\ifx \@tempc \@empty \let \@tempc \@tempb \let \@tempb \@tempa \fi \ifx
  \@tempb \@empty \def\@tempb {arXiv}\fi \@ifundefined
  {mn@eprint@\@tempb}{\@tempb:\@tempc}{\expandafter \expandafter \csname
  mn@eprint@\@tempb\endcsname \expandafter{\@tempc}}}

\bibitem[\protect\citeauthoryear{{Beck}}{{Beck}}{2016}]{2016A&ARv..24....4B}
{Beck} R.,  2016, \mn@doi [\aapr] {10.1007/s00159-015-0084-4}, \href
  {http://adsabs.harvard.edu/abs/2016A\&ARv..24....4B} {24, 4}

\bibitem[\protect\citeauthoryear{{Becker}, {Bolton}, {Haehnelt}  \&
  {Sargent}}{{Becker} et~al.}{2011}]{2011MNRAS.410.1096B}
{Becker} G.~D.,  {Bolton} J.~S.,  {Haehnelt} M.~G.,   {Sargent} W.~L.~W.,
  2011, \mn@doi [\mnras] {10.1111/j.1365-2966.2010.17507.x}, \href
  {http://adsabs.harvard.edu/abs/2011MNRAS.410.1096B} {410, 1096}

\bibitem[\protect\citeauthoryear{{Bell}}{{Bell}}{2015}]{2015MNRAS.447.2224B}
{Bell} A.~R.,  2015, \mn@doi [\mnras] {10.1093/mnras/stu2596}, \href
  {http://adsabs.harvard.edu/abs/2015MNRAS.447.2224B} {447, 2224}

\bibitem[\protect\citeauthoryear{{Bernet}, {Miniati}, {Lilly}, {Kronberg}  \&
  {Dessauges-Zavadsky}}{{Bernet} et~al.}{2008}]{2008Natur.454..302B}
{Bernet} M.~L.,  {Miniati} F.,  {Lilly} S.~J.,  {Kronberg} P.~P.,
  {Dessauges-Zavadsky} M.,  2008, \mn@doi [\nat] {10.1038/nature07105}, \href
  {http://adsabs.harvard.edu/abs/2008Natur.454..302B} {454, 302}

\bibitem[\protect\citeauthoryear{{Bertone}, {Vogt}  \& {En{\ss}lin}}{{Bertone}
  et~al.}{2006}]{2006MNRAS.370..319B}
{Bertone} S.,  {Vogt} C.,   {En{\ss}lin} T.,  2006, \mn@doi [\mnras]
  {10.1111/j.1365-2966.2006.10474.x}, \href
  {http://adsabs.harvard.edu/abs/2006MNRAS.370..319B} {370, 319}

\bibitem[\protect\citeauthoryear{{Bhat} \& {Subramanian}}{{Bhat} \&
  {Subramanian}}{2015}]{2015JPlPh..81e3902B}
{Bhat} P.,  {Subramanian} K.,  2015, \mn@doi [Journal of Plasma Physics]
  {10.1017/S0022377815000616}, \href
  {http://adsabs.harvard.edu/abs/2015JPlPh..81e3902B} {81, 395810502}

\bibitem[\protect\citeauthoryear{{Boera}, {Murphy}, {Becker}  \&
  {Bolton}}{{Boera} et~al.}{2014}]{2014MNRAS.441.1916B}
{Boera} E.,  {Murphy} M.~T.,  {Becker} G.~D.,   {Bolton} J.~S.,  2014, \mn@doi
  [\mnras] {10.1093/mnras/stu660}, \href
  {http://adsabs.harvard.edu/abs/2014MNRAS.441.1916B} {441, 1916}

\bibitem[\protect\citeauthoryear{{Bolton}, {Oh}  \& {Furlanetto}}{{Bolton}
  et~al.}{2009}]{2009MNRAS.395..736B}
{Bolton} J.~S.,  {Oh} S.~P.,   {Furlanetto} S.~R.,  2009, \mn@doi [\mnras]
  {10.1111/j.1365-2966.2009.14597.x}, \href
  {http://adsabs.harvard.edu/abs/2009MNRAS.395..736B} {395, 736}

\bibitem[\protect\citeauthoryear{{Brandenburg} \& {Subramanian}}{{Brandenburg}
  \& {Subramanian}}{2005}]{2005PhR...417....1B}
{Brandenburg} A.,  {Subramanian} K.,  2005, \mn@doi [\physrep]
  {10.1016/j.physrep.2005.06.005}, \href
  {http://adsabs.harvard.edu/abs/2005PhR...417....1B} {417, 1}

\bibitem[\protect\citeauthoryear{{Breitschwerdt}, {McKenzie}  \&
  {Voelk}}{{Breitschwerdt} et~al.}{1991}]{1991A&A...245...79B}
{Breitschwerdt} D.,  {McKenzie} J.~F.,   {Voelk} H.~J.,  1991, \aap, \href
  {http://adsabs.harvard.edu/abs/1991A%26A...245...79B} {245, 79}

\bibitem[\protect\citeauthoryear{{Bromm} \& {Loeb}}{{Bromm} \&
  {Loeb}}{2002}]{2002ApJ...575..111B}
{Bromm} V.,  {Loeb} A.,  2002, \mn@doi [\apj] {10.1086/341189}, \href
  {http://adsabs.harvard.edu/abs/2002ApJ...575..111B} {575, 111}

\bibitem[\protect\citeauthoryear{{Carswell}, {Schaye}  \& {Kim}}{{Carswell}
  et~al.}{2002}]{2002ApJ...578...43C}
{Carswell} B.,  {Schaye} J.,   {Kim} T.-S.,  2002, \mn@doi [\apj]
  {10.1086/342404}, \href {http://adsabs.harvard.edu/abs/2002ApJ...578...43C}
  {578, 43}

\bibitem[\protect\citeauthoryear{{Chen}, {Helsby}, {Gauthier}, {Shectman},
  {Thompson}  \& {Tinker}}{{Chen} et~al.}{2010}]{2010ApJ...714.1521C}
{Chen} H.-W.,  {Helsby} J.~E.,  {Gauthier} J.-R.,  {Shectman} S.~A.,
  {Thompson} I.~B.,   {Tinker} J.~L.,  2010, \mn@doi [\apj]
  {10.1088/0004-637X/714/2/1521}, \href
  {http://adsabs.harvard.edu/abs/2010ApJ...714.1521C} {714, 1521}

\bibitem[\protect\citeauthoryear{{Chevalier}}{{Chevalier}}{1983}]{1983ApJ...272..765C}
{Chevalier} R.~A.,  1983, \mn@doi [\apj] {10.1086/161338}, \href
  {http://adsabs.harvard.edu/abs/1983ApJ...272..765C} {272, 765}

\bibitem[\protect\citeauthoryear{{Chevalier} \& {Clegg}}{{Chevalier} \&
  {Clegg}}{1985}]{1985Natur.317...44C}
{Chevalier} R.~A.,  {Clegg} A.~W.,  1985, \mn@doi [\nat] {10.1038/317044a0},
  \href {http://adsabs.harvard.edu/abs/1985Natur.317...44C} {317, 44}

\bibitem[\protect\citeauthoryear{{Chiu} \& {Ostriker}}{{Chiu} \&
  {Ostriker}}{2000}]{2000ApJ...534..507C}
{Chiu} W.~A.,  {Ostriker} J.~P.,  2000, \mn@doi [\apj] {10.1086/308780}, \href
  {http://adsabs.harvard.edu/abs/2000ApJ...534..507C} {534, 507}

\bibitem[\protect\citeauthoryear{{Choudhury} \& {Srianand}}{{Choudhury} \&
  {Srianand}}{2002}]{2002MNRAS.336L..27R}
{Choudhury} T.~R.,  {Srianand} R.,  2002, \mn@doi [\mnras]
  {10.1046/j.1365-8711.2002.05984.x}, \href
  {http://adsabs.harvard.edu/abs/2002MNRAS.336L..27R} {336, L27}

\bibitem[\protect\citeauthoryear{{Cooksey}, {Kao}, {Simcoe}, {O'Meara}  \&
  {Prochaska}}{{Cooksey} et~al.}{2013}]{2013ApJ...763...37C}
{Cooksey} K.~L.,  {Kao} M.~M.,  {Simcoe} R.~A.,  {O'Meara} J.~M.,   {Prochaska}
  J.~X.,  2013, \mn@doi [\apj] {10.1088/0004-637X/763/1/37}, \href
  {http://adsabs.harvard.edu/abs/2013ApJ...763...37C} {763, 37}

\bibitem[\protect\citeauthoryear{{D'Odorico} et~al.,}{{D'Odorico}
  et~al.}{2013}]{2013MNRAS.435.1198D}
{D'Odorico} V.,  et~al., 2013, \mn@doi [\mnras] {10.1093/mnras/stt1365}, \href
  {http://adsabs.harvard.edu/abs/2013MNRAS.435.1198D} {435, 1198}

\bibitem[\protect\citeauthoryear{{D'Odorico} et~al.,}{{D'Odorico}
  et~al.}{2016}]{2016MNRAS.463.2690D}
{D'Odorico} V.,  et~al., 2016, \mn@doi [\mnras] {10.1093/mnras/stw2161}, \href
  {http://adsabs.harvard.edu/abs/2016MNRAS.463.2690D} {463, 2690}

\bibitem[\protect\citeauthoryear{{Dav{\'e}}, {Oppenheimer}  \&
  {Sivanandam}}{{Dav{\'e}} et~al.}{2008}]{2008MNRAS.391..110D}
{Dav{\'e}} R.,  {Oppenheimer} B.~D.,   {Sivanandam} S.,  2008, \mn@doi [\mnras]
  {10.1111/j.1365-2966.2008.13906.x}, \href
  {http://adsabs.harvard.edu/abs/2008MNRAS.391..110D} {391, 110}

\bibitem[\protect\citeauthoryear{{Dav{\'e}}, {Oppenheimer}  \&
  {Finlator}}{{Dav{\'e}} et~al.}{2011}]{2011MNRAS.415...11D}
{Dav{\'e}} R.,  {Oppenheimer} B.~D.,   {Finlator} K.,  2011, \mn@doi [\mnras]
  {10.1111/j.1365-2966.2011.18680.x}, \href
  {http://adsabs.harvard.edu/abs/2011MNRAS.415...11D} {415, 11}

\bibitem[\protect\citeauthoryear{{Dekel} \& {Mandelker}}{{Dekel} \&
  {Mandelker}}{2014}]{2014MNRAS.444.2071D}
{Dekel} A.,  {Mandelker} N.,  2014, \mn@doi [\mnras] {10.1093/mnras/stu1427},
  \href {http://adsabs.harvard.edu/abs/2014MNRAS.444.2071D} {444, 2071}

\bibitem[\protect\citeauthoryear{{Dekel}, {Zolotov}, {Tweed}, {Cacciato},
  {Ceverino}  \& {Primack}}{{Dekel} et~al.}{2013}]{2013MNRAS.435..999D}
{Dekel} A.,  {Zolotov} A.,  {Tweed} D.,  {Cacciato} M.,  {Ceverino} D.,
  {Primack} J.~R.,  2013, \mn@doi [\mnras] {10.1093/mnras/stt1338}, \href
  {http://adsabs.harvard.edu/abs/2013MNRAS.435..999D} {435, 999}

\bibitem[\protect\citeauthoryear{{Durrer} \& {Neronov}}{{Durrer} \&
  {Neronov}}{2013}]{2013A&ARv..21...62D}
{Durrer} R.,  {Neronov} A.,  2013, \mn@doi [\aapr] {10.1007/s00159-013-0062-7},
  \href {http://adsabs.harvard.edu/abs/2013A\&ARv..21...62D} {21, 62}

\bibitem[\protect\citeauthoryear{{Farnes}, {O'Sullivan}, {Corrigan}  \&
  {Gaensler}}{{Farnes} et~al.}{2014}]{2014ApJ...795...63F}
{Farnes} J.~S.,  {O'Sullivan} S.~P.,  {Corrigan} M.~E.,   {Gaensler} B.~M.,
  2014, \mn@doi [\apj] {10.1088/0004-637X/795/1/63}, \href
  {http://adsabs.harvard.edu/abs/2014ApJ...795...63F} {795, 63}

\bibitem[\protect\citeauthoryear{{Furlanetto} \& {Loeb}}{{Furlanetto} \&
  {Loeb}}{2001}]{2001ApJ...556..619F}
{Furlanetto} S.~R.,  {Loeb} A.,  2001, \mn@doi [\apj] {10.1086/321630}, \href
  {http://adsabs.harvard.edu/abs/2001ApJ...556..619F} {556, 619}

\bibitem[\protect\citeauthoryear{{Furlanetto} \& {Loeb}}{{Furlanetto} \&
  {Loeb}}{2003}]{2003ApJ...588...18F}
{Furlanetto} S.~R.,  {Loeb} A.,  2003, \mn@doi [\apj] {10.1086/374045}, \href
  {http://adsabs.harvard.edu/abs/2003ApJ...588...18F} {588, 18}

\bibitem[\protect\citeauthoryear{{Furlanetto} \& {Oh}}{{Furlanetto} \&
  {Oh}}{2008}]{2008ApJ...682...14F}
{Furlanetto} S.~R.,  {Oh} S.~P.,  2008, \mn@doi [\apj] {10.1086/589613}, \href
  {http://adsabs.harvard.edu/abs/2008ApJ...682...14F} {682, 14}

\bibitem[\protect\citeauthoryear{{Girichidis} et~al.,}{{Girichidis}
  et~al.}{2016}]{2016ApJ...816L..19G}
{Girichidis} P.,  et~al., 2016, \mn@doi [\apjl] {10.3847/2041-8205/816/2/L19},
  \href {http://adsabs.harvard.edu/abs/2016ApJ...816L..19G} {816, L19}

\bibitem[\protect\citeauthoryear{{Haiman}, {Abel}  \& {Rees}}{{Haiman}
  et~al.}{2000}]{2000ApJ...534...11H}
{Haiman} Z.,  {Abel} T.,   {Rees} M.~J.,  2000, \mn@doi [\apj]
  {10.1086/308723}, \href {http://adsabs.harvard.edu/abs/2000ApJ...534...11H}
  {534, 11}

\bibitem[\protect\citeauthoryear{{Ipavich}}{{Ipavich}}{1975}]{1975ApJ...196..107I}
{Ipavich} F.~M.,  1975, \mn@doi [\apj] {10.1086/153397}, \href
  {http://adsabs.harvard.edu/abs/1975ApJ...196..107I} {196, 107}

\bibitem[\protect\citeauthoryear{{Jose}, {Subramanian}, {Srianand}  \&
  {Samui}}{{Jose} et~al.}{2013}]{2013MNRAS.429.2333J}
{Jose} C.,  {Subramanian} K.,  {Srianand} R.,   {Samui} S.,  2013, \mn@doi
  [\mnras] {10.1093/mnras/sts503}, \href
  {http://adsabs.harvard.edu/abs/2013MNRAS.429.2333J} {429, 2333}

\bibitem[\protect\citeauthoryear{{Jose}, {Srianand}  \& {Subramanian}}{{Jose}
  et~al.}{2014}]{2014MNRAS.443.3341J}
{Jose} C.,  {Srianand} R.,   {Subramanian} K.,  2014, \mn@doi [\mnras]
  {10.1093/mnras/stu1339}, \href
  {http://adsabs.harvard.edu/abs/2014MNRAS.443.3341J} {443, 3341}

\bibitem[\protect\citeauthoryear{{Jubelgas}, {Springel}, {En{\ss}lin}  \&
  {Pfrommer}}{{Jubelgas} et~al.}{2008}]{2008A&A...481...33J}
{Jubelgas} M.,  {Springel} V.,  {En{\ss}lin} T.,   {Pfrommer} C.,  2008,
  \mn@doi [\aap] {10.1051/0004-6361:20065295}, \href
  {http://adsabs.harvard.edu/abs/2008A\&A...481...33J} {481, 33}

\bibitem[\protect\citeauthoryear{{Kang} \& {Jones}}{{Kang} \&
  {Jones}}{2003}]{2003ICRC....4.2039K}
{Kang} H.,  {Jones} T.~W.,  2003, International Cosmic Ray Conference, \href
  {http://adsabs.harvard.edu/abs/2003ICRC....4.2039K} {4, 2039}

\bibitem[\protect\citeauthoryear{{Kang} \& {Jones}}{{Kang} \&
  {Jones}}{2005}]{2005ApJ...620...44K}
{Kang} H.,  {Jones} T.~W.,  2005, \mn@doi [\apj] {10.1086/426855}, \href
  {http://adsabs.harvard.edu/abs/2005ApJ...620...44K} {620, 44}

\bibitem[\protect\citeauthoryear{Kazantsev}{Kazantsev}{1967}]{Kaz68}
Kazantsev A.~P.,  1967, Soviet Journal of Experimental and Theoretical Physics,
  53, 1807

\bibitem[\protect\citeauthoryear{{Kulsrud}}{{Kulsrud}}{2004}]{kulsrud_book}
{Kulsrud} R.~M.,  2004, Plasma Physics for Astrophysics.
Princeton Series in Astrophysics, Princeton Univ. Press

\bibitem[\protect\citeauthoryear{{Kulsrud} \& {Cesarsky}}{{Kulsrud} \&
  {Cesarsky}}{1971}]{1971ApL.....8..189K}
{Kulsrud} R.~M.,  {Cesarsky} C.~J.,  1971, \aplett, \href
  {http://adsabs.harvard.edu/abs/1971ApL.....8..189K} {8, 189}

\bibitem[\protect\citeauthoryear{{Kulsrud} \& {Pearce}}{{Kulsrud} \&
  {Pearce}}{1969}]{1969ApJ...156..445K}
{Kulsrud} R.,  {Pearce} W.~P.,  1969, \mn@doi [\apj] {10.1086/149981}, \href
  {http://adsabs.harvard.edu/abs/1969ApJ...156..445K} {156, 445}

\bibitem[\protect\citeauthoryear{{Kulsrud} \& {Zweibel}}{{Kulsrud} \&
  {Zweibel}}{2008}]{2008RPPh...71d6901K}
{Kulsrud} R.~M.,  {Zweibel} E.~G.,  2008, \mn@doi [Reports on Progress in
  Physics] {10.1088/0034-4885/71/4/046901}, \href
  {http://adsabs.harvard.edu/abs/2008RPPh...71d6901K} {71, 046901}

\bibitem[\protect\citeauthoryear{{Madau}, {Ferrara}  \& {Rees}}{{Madau}
  et~al.}{2001}]{2001ApJ...555...92M}
{Madau} P.,  {Ferrara} A.,   {Rees} M.~J.,  2001, \mn@doi [\apj]
  {10.1086/321474}, \href {http://adsabs.harvard.edu/abs/2001ApJ...555...92M}
  {555, 92}

\bibitem[\protect\citeauthoryear{{Makino}, {Sasaki}  \& {Suto}}{{Makino}
  et~al.}{1998}]{1998ApJ...497..555M}
{Makino} N.,  {Sasaki} S.,   {Suto} Y.,  1998, \mn@doi [\apj] {10.1086/305507},
  \href {http://adsabs.harvard.edu/abs/1998ApJ...497..555M} {497, 555}

\bibitem[\protect\citeauthoryear{{Martin}, {Shapley}, {Coil}, {Kornei},
  {Bundy}, {Weiner}, {Noeske}  \& {Schiminovich}}{{Martin}
  et~al.}{2012}]{2012ApJ...760..127M}
{Martin} C.~L.,  {Shapley} A.~E.,  {Coil} A.~L.,  {Kornei} K.~A.,  {Bundy} K.,
  {Weiner} B.~J.,  {Noeske} K.~G.,   {Schiminovich} D.,  2012, \mn@doi [\apj]
  {10.1088/0004-637X/760/2/127}, \href
  {http://adsabs.harvard.edu/abs/2012ApJ...760..127M} {760, 127}

\bibitem[\protect\citeauthoryear{{Martin}, {Shapley}, {Coil}, {Kornei},
  {Murray}  \& {Pancoast}}{{Martin} et~al.}{2013}]{2013ApJ...770...41M}
{Martin} C.~L.,  {Shapley} A.~E.,  {Coil} A.~L.,  {Kornei} K.~A.,  {Murray} N.,
    {Pancoast} A.,  2013, \mn@doi [\apj] {10.1088/0004-637X/770/1/41}, \href
  {http://adsabs.harvard.edu/abs/2013ApJ...770...41M} {770, 41}

\bibitem[\protect\citeauthoryear{{Mori}, {Ferrara}  \& {Madau}}{{Mori}
  et~al.}{2002}]{2002ApJ...571...40M}
{Mori} M.,  {Ferrara} A.,   {Madau} P.,  2002, \mn@doi [\apj] {10.1086/339913},
  \href {http://adsabs.harvard.edu/abs/2002ApJ...571...40M} {571, 40}

\bibitem[\protect\citeauthoryear{{Muzahid}, {Srianand}, {Bergeron}  \&
  {Petitjean}}{{Muzahid} et~al.}{2012}]{2012MNRAS.421..446M}
{Muzahid} S.,  {Srianand} R.,  {Bergeron} J.,   {Petitjean} P.,  2012, \mn@doi
  [\mnras] {10.1111/j.1365-2966.2011.20324.x}, \href
  {http://adsabs.harvard.edu/abs/2012MNRAS.421..446M} {421, 446}

\bibitem[\protect\citeauthoryear{{Naab} \& {Ostriker}}{{Naab} \&
  {Ostriker}}{2017}]{2017ARA&A..55...59N}
{Naab} T.,  {Ostriker} J.~P.,  2017, \mn@doi [\araa]
  {10.1146/annurev-astro-081913-040019}, \href
  {http://adsabs.harvard.edu/abs/2017ARA%26A..55...59N} {55, 59}

\bibitem[\protect\citeauthoryear{{Navarro}, {Frenk}  \& {White}}{{Navarro}
  et~al.}{1997}]{1997ApJ...490..493N}
{Navarro} J.~F.,  {Frenk} C.~S.,   {White} S.~D.~M.,  1997, \apj, \href
  {http://adsabs.harvard.edu/abs/1997ApJ...490..493N} {490, 493}

\bibitem[\protect\citeauthoryear{{Neronov} \& {Vovk}}{{Neronov} \&
  {Vovk}}{2010}]{2010Sci...328...73N}
{Neronov} A.,  {Vovk} I.,  2010, \mn@doi [Science] {10.1126/science.1184192},
  \href {http://adsabs.harvard.edu/abs/2010Sci...328...73N} {328, 73}

\bibitem[\protect\citeauthoryear{{Nielsen}, {Churchill}, {Kacprzak}  \&
  {Murphy}}{{Nielsen} et~al.}{2013}]{2013ApJ...776..114N}
{Nielsen} N.~M.,  {Churchill} C.~W.,  {Kacprzak} G.~G.,   {Murphy} M.~T.,
  2013, \mn@doi [\apj] {10.1088/0004-637X/776/2/114}, \href
  {http://adsabs.harvard.edu/abs/2013ApJ...776..114N} {776, 114}

\bibitem[\protect\citeauthoryear{{Ostriker} \& {McKee}}{{Ostriker} \&
  {McKee}}{1988}]{1988RvMP...60....1O}
{Ostriker} J.~P.,  {McKee} C.~F.,  1988, \mn@doi [Reviews of Modern Physics]
  {10.1103/RevModPhys.60.1}, \href
  {http://adsabs.harvard.edu/abs/1988RvMP...60....1O} {60, 1}

\bibitem[\protect\citeauthoryear{{Pakmor}, {Pfrommer}, {Simpson}  \&
  {Springel}}{{Pakmor} et~al.}{2016}]{2016ApJ...824L..30P}
{Pakmor} R.,  {Pfrommer} C.,  {Simpson} C.~M.,   {Springel} V.,  2016, \mn@doi
  [\apjl] {10.3847/2041-8205/824/2/L30}, \href
  {http://adsabs.harvard.edu/abs/2016ApJ...824L..30P} {824, L30}

\bibitem[\protect\citeauthoryear{{Pfrommer}, {Pakmor}, {Schaal}, {Simpson}  \&
  {Springel}}{{Pfrommer} et~al.}{2017}]{2017MNRAS.465.4500P}
{Pfrommer} C.,  {Pakmor} R.,  {Schaal} K.,  {Simpson} C.~M.,   {Springel} V.,
  2017, \mn@doi [\mnras] {10.1093/mnras/stw2941}, \href
  {http://adsabs.harvard.edu/abs/2017MNRAS.465.4500P} {465, 4500}

\bibitem[\protect\citeauthoryear{{Planck Collaboration} et~al.,}{{Planck
  Collaboration} et~al.}{2016a}]{2016A&A...594A..13P}
{Planck Collaboration} et~al., 2016a, \mn@doi [\aap]
  {10.1051/0004-6361/201525830}, \href
  {http://adsabs.harvard.edu/abs/2016A%26A...594A..13P} {594, A13}

\bibitem[\protect\citeauthoryear{{Planck Collaboration} et~al.,}{{Planck
  Collaboration} et~al.}{2016b}]{2016A&A...596A.108P}
{Planck Collaboration} et~al., 2016b, \mn@doi [\aap]
  {10.1051/0004-6361/201628897}, \href
  {http://adsabs.harvard.edu/abs/2016A%26A...596A.108P} {596, A108}

\bibitem[\protect\citeauthoryear{{Press} \& {Schechter}}{{Press} \&
  {Schechter}}{1974}]{1974ApJ...187..425P}
{Press} W.~H.,  {Schechter} P.,  1974, \mn@doi [\apj] {10.1086/152650}, \href
  {http://adsabs.harvard.edu/abs/1974ApJ...187..425P} {187, 425}

\bibitem[\protect\citeauthoryear{{Prochaska}, {Weiner}, {Chen}, {Mulchaey}  \&
  {Cooksey}}{{Prochaska} et~al.}{2011}]{2011ApJ...740...91P}
{Prochaska} J.~X.,  {Weiner} B.,  {Chen} H.-W.,  {Mulchaey} J.,   {Cooksey} K.,
   2011, \mn@doi [\apj] {10.1088/0004-637X/740/2/91}, \href
  {http://adsabs.harvard.edu/abs/2011ApJ...740...91P} {740, 91}

\bibitem[\protect\citeauthoryear{{Prochaska}, {Lau}  \& {Hennawi}}{{Prochaska}
  et~al.}{2014}]{2014ApJ...796..140P}
{Prochaska} J.~X.,  {Lau} M.~W.,   {Hennawi} J.~F.,  2014, \mn@doi [\apj]
  {10.1088/0004-637X/796/2/140}, \href
  {http://adsabs.harvard.edu/abs/2014ApJ...796..140P} {796, 140}

\bibitem[\protect\citeauthoryear{{Ricotti}, {Gnedin}  \& {Shull}}{{Ricotti}
  et~al.}{2000}]{2000ApJ...534...41R}
{Ricotti} M.,  {Gnedin} N.~Y.,   {Shull} J.~M.,  2000, \mn@doi [\apj]
  {10.1086/308733}, \href {http://adsabs.harvard.edu/abs/2000ApJ...534...41R}
  {534, 41}

\bibitem[\protect\citeauthoryear{{Rodrigues}, {Shukurov}, {Fletcher}  \&
  {Baugh}}{{Rodrigues} et~al.}{2015}]{2015MNRAS.450.3472R}
{Rodrigues} L.~F.~S.,  {Shukurov} A.,  {Fletcher} A.,   {Baugh} C.~M.,  2015,
  \mn@doi [\mnras] {10.1093/mnras/stv816}, \href
  {http://adsabs.harvard.edu/abs/2015MNRAS.450.3472R} {450, 3472}

\bibitem[\protect\citeauthoryear{{Rubin}, {Prochaska}, {Koo}, {Phillips},
  {Martin}  \& {Winstrom}}{{Rubin} et~al.}{2014}]{2014ApJ...794..156R}
{Rubin} K.~H.~R.,  {Prochaska} J.~X.,  {Koo} D.~C.,  {Phillips} A.~C.,
  {Martin} C.~L.,   {Winstrom} L.~O.,  2014, \mn@doi [\apj]
  {10.1088/0004-637X/794/2/156}, \href
  {http://adsabs.harvard.edu/abs/2014ApJ...794..156R} {794, 156}

\bibitem[\protect\citeauthoryear{{Ruszkowski}, {Yang}  \&
  {Zweibel}}{{Ruszkowski} et~al.}{2017}]{2017ApJ...834..208R}
{Ruszkowski} M.,  {Yang} H.-Y.~K.,   {Zweibel} E.,  2017, \mn@doi [\apj]
  {10.3847/1538-4357/834/2/208}, \href
  {http://adsabs.harvard.edu/abs/2017ApJ...834..208R} {834, 208}

\bibitem[\protect\citeauthoryear{{Ruzmaikin}, {Shukurov}  \&
  {Sokoloff}}{{Ruzmaikin} et~al.}{1988}]{RSS88}
{Ruzmaikin} A.~A.,  {Shukurov} A.~M.,   {Sokoloff} D.~D.,  1988, Magnetic
  Fields of Galaxies.
Kluwer: Dordrecht

\bibitem[\protect\citeauthoryear{{Ryan-Weber}, {Pettini}  \&
  {Madau}}{{Ryan-Weber} et~al.}{2006}]{2006MNRAS.371L..78R}
{Ryan-Weber} E.~V.,  {Pettini} M.,   {Madau} P.,  2006, \mn@doi [\mnras]
  {10.1111/j.1745-3933.2006.00212.x}, \href
  {http://adsabs.harvard.edu/abs/2006MNRAS.371L..78R} {371, L78}

\bibitem[\protect\citeauthoryear{{Salem} \& {Bryan}}{{Salem} \&
  {Bryan}}{2014}]{2014MNRAS.437.3312S}
{Salem} M.,  {Bryan} G.~L.,  2014, \mn@doi [\mnras] {10.1093/mnras/stt2121},
  \href {http://adsabs.harvard.edu/abs/2014MNRAS.437.3312S} {437, 3312}

\bibitem[\protect\citeauthoryear{{Samui}}{{Samui}}{2014}]{2014NewA...30...89S}
{Samui} S.,  2014, \mn@doi [\na] {10.1016/j.newast.2014.01.010}, \href
  {http://adsabs.harvard.edu/abs/2014NewA...30...89S} {30, 89}

\bibitem[\protect\citeauthoryear{{Samui}, {Subramanian}  \& {Srianand}}{{Samui}
  et~al.}{2005}]{2005ICRC....9..215S}
{Samui} S.,  {Subramanian} K.,   {Srianand} R.,  2005, International Cosmic Ray
  Conference, \href {http://adsabs.harvard.edu/abs/2005ICRC....9..215S} {9,
  215}

\bibitem[\protect\citeauthoryear{{Samui}, {Srianand}  \& {Subramanian}}{{Samui}
  et~al.}{2007}]{2007MNRAS.377..285S}
{Samui} S.,  {Srianand} R.,   {Subramanian} K.,  2007, \mn@doi [\mnras]
  {10.1111/j.1365-2966.2007.11603.x}, \href
  {http://adsabs.harvard.edu/abs/2007MNRAS.377..285S} {377, 285}

\bibitem[\protect\citeauthoryear{{Samui}, {Subramanian}  \& {Srianand}}{{Samui}
  et~al.}{2008}]{2008MNRAS.385..783S}
{Samui} S.,  {Subramanian} K.,   {Srianand} R.,  2008, \mn@doi [\mnras]
  {10.1111/j.1365-2966.2008.12932.x}, \href
  {http://adsabs.harvard.edu/abs/2008MNRAS.385..783S} {385, 783}

\bibitem[\protect\citeauthoryear{{Samui}, {Subramanian}  \& {Srianand}}{{Samui}
  et~al.}{2009a}]{2009NewA...14..591S}
{Samui} S.,  {Subramanian} K.,   {Srianand} R.,  2009a, \mn@doi [\na]
  {10.1016/j.newast.2009.02.006}, \href
  {http://adsabs.harvard.edu/abs/2009NewA...14..591S} {14, 591}

\bibitem[\protect\citeauthoryear{{Samui}, {Srianand}  \& {Subramanian}}{{Samui}
  et~al.}{2009b}]{2009MNRAS.398.2061S}
{Samui} S.,  {Srianand} R.,   {Subramanian} K.,  2009b, \mn@doi [\mnras]
  {10.1111/j.1365-2966.2009.15245.x}, \href
  {http://adsabs.harvard.edu/abs/2009MNRAS.398.2061S} {398, 2061}

\bibitem[\protect\citeauthoryear{{Samui}, {Subramanian}  \& {Srianand}}{{Samui}
  et~al.}{2010}]{2010MNRAS.402.2778S}
{Samui} S.,  {Subramanian} K.,   {Srianand} R.,  2010, \mn@doi [\mnras]
  {10.1111/j.1365-2966.2009.16099.x}, \href
  {http://adsabs.harvard.edu/abs/2010MNRAS.402.2778S} {402, 2778}

\bibitem[\protect\citeauthoryear{{Sasaki}}{{Sasaki}}{1994}]{1994PASJ...46..427S}
{Sasaki} S.,  1994, \pasj, \href
  {http://adsabs.harvard.edu/abs/1994PASJ...46..427S} {46, 427}

\bibitem[\protect\citeauthoryear{{Scannapieco}, {Ferrara}  \&
  {Madau}}{{Scannapieco} et~al.}{2002}]{2002ApJ...574..590S}
{Scannapieco} E.,  {Ferrara} A.,   {Madau} P.,  2002, \mn@doi [\apj]
  {10.1086/341114}, \href {http://adsabs.harvard.edu/abs/2002ApJ...574..590S}
  {574, 590}

\bibitem[\protect\citeauthoryear{{Scannapieco}, {Tissera}, {White}  \&
  {Springel}}{{Scannapieco} et~al.}{2005}]{2005MNRAS.364..552S}
{Scannapieco} C.,  {Tissera} P.~B.,  {White} S.~D.~M.,   {Springel} V.,  2005,
  \mn@doi [\mnras] {10.1111/j.1365-2966.2005.09574.x}, \href
  {http://adsabs.harvard.edu/abs/2005MNRAS.364..552S} {364, 552}

\bibitem[\protect\citeauthoryear{{Scannapieco}, {Tissera}, {White}  \&
  {Springel}}{{Scannapieco} et~al.}{2006}]{2006MNRAS.371.1125S}
{Scannapieco} C.,  {Tissera} P.~B.,  {White} S.~D.~M.,   {Springel} V.,  2006,
  \mn@doi [\mnras] {10.1111/j.1365-2966.2006.10785.x}, \href
  {http://adsabs.harvard.edu/abs/2006MNRAS.371.1125S} {371, 1125}

\bibitem[\protect\citeauthoryear{{Scannapieco} et~al.,}{{Scannapieco}
  et~al.}{2012}]{2012MNRAS.423.1726S}
{Scannapieco} C.,  et~al., 2012, \mn@doi [\mnras]
  {10.1111/j.1365-2966.2012.20993.x}, \href
  {http://adsabs.harvard.edu/abs/2012MNRAS.423.1726S} {423, 1726}

\bibitem[\protect\citeauthoryear{{Schaye}}{{Schaye}}{2001}]{2001ApJ...559..507S}
{Schaye} J.,  2001, \mn@doi [\apj] {10.1086/322421}, \href
  {http://adsabs.harvard.edu/abs/2001ApJ...559..507S} {559, 507}

\bibitem[\protect\citeauthoryear{{Schaye}, {Theuns}, {Rauch}, {Efstathiou}  \&
  {Sargent}}{{Schaye} et~al.}{2000}]{2000MNRAS.318..817S}
{Schaye} J.,  {Theuns} T.,  {Rauch} M.,  {Efstathiou} G.,   {Sargent} W.~L.~W.,
   2000, \mn@doi [\mnras] {10.1046/j.1365-8711.2000.03815.x}, \href
  {http://adsabs.harvard.edu/abs/2000MNRAS.318..817S} {318, 817}

\bibitem[\protect\citeauthoryear{{Sharma} \& {Nath}}{{Sharma} \&
  {Nath}}{2013}]{2013ApJ...763...17S}
{Sharma} M.,  {Nath} B.~B.,  2013, \mn@doi [\apj] {10.1088/0004-637X/763/1/17},
  \href {http://adsabs.harvard.edu/abs/2013ApJ...763...17S} {763, 17}

\bibitem[\protect\citeauthoryear{{Sharma}, {Roy}, {Nath}  \&
  {Shchekinov}}{{Sharma} et~al.}{2014}]{2014MNRAS.443.3463S}
{Sharma} P.,  {Roy} A.,  {Nath} B.~B.,   {Shchekinov} Y.,  2014, \mn@doi
  [\mnras] {10.1093/mnras/stu1307}, \href
  {http://adsabs.harvard.edu/abs/2014MNRAS.443.3463S} {443, 3463}

\bibitem[\protect\citeauthoryear{{Sheth} \& {Tormen}}{{Sheth} \&
  {Tormen}}{1999}]{1999MNRAS.308..119S}
{Sheth} R.~K.,  {Tormen} G.,  1999, \mn@doi [\mnras]
  {10.1046/j.1365-8711.1999.02692.x}, \href
  {http://adsabs.harvard.edu/abs/1999MNRAS.308..119S} {308, 119}

\bibitem[\protect\citeauthoryear{{Songaila} \& {Cowie}}{{Songaila} \&
  {Cowie}}{1996}]{1996AJ....112..335S}
{Songaila} A.,  {Cowie} L.~L.,  1996, \mn@doi [\aj] {10.1086/118018}, \href
  {http://adsabs.harvard.edu/abs/1996AJ....112..335S} {112, 335}

\bibitem[\protect\citeauthoryear{{Subramanian}}{{Subramanian}}{2016}]{2016RPPh...79g6901S}
{Subramanian} K.,  2016, \mn@doi [Reports on Progress in Physics]
  {10.1088/0034-4885/79/7/076901}, \href
  {http://adsabs.harvard.edu/abs/2016RPPh...79g6901S} {79, 076901}

\bibitem[\protect\citeauthoryear{{Tavecchio}, {Ghisellini}, {Bonnoli}  \&
  {Foschini}}{{Tavecchio} et~al.}{2011}]{2011MNRAS.414.3566T}
{Tavecchio} F.,  {Ghisellini} G.,  {Bonnoli} G.,   {Foschini} L.,  2011,
  \mn@doi [\mnras] {10.1111/j.1365-2966.2011.18657.x}, \href
  {http://adsabs.harvard.edu/abs/2011MNRAS.414.3566T} {414, 3566}

\bibitem[\protect\citeauthoryear{{Tegmark}, {Silk}  \& {Evrard}}{{Tegmark}
  et~al.}{1993}]{1993ApJ...417...54T}
{Tegmark} M.,  {Silk} J.,   {Evrard} A.,  1993, \mn@doi [\apj]
  {10.1086/173290}, \href {http://adsabs.harvard.edu/abs/1993ApJ...417...54T}
  {417, 54}

\bibitem[\protect\citeauthoryear{{Tegmark}, {Silk}, {Rees}, {Blanchard}, {Abel}
   \& {Palla}}{{Tegmark} et~al.}{1997}]{1997ApJ...474....1T}
{Tegmark} M.,  {Silk} J.,  {Rees} M.~J.,  {Blanchard} A.,  {Abel} T.,   {Palla}
  F.,  1997, \mn@doi [\apj] {10.1086/303434}, \href
  {http://adsabs.harvard.edu/abs/1997ApJ...474....1T} {474, 1}

\bibitem[\protect\citeauthoryear{{Visbal}, {Bryan}  \& {Haiman}}{{Visbal}
  et~al.}{2017}]{2017MNRAS.469.1456V}
{Visbal} E.,  {Bryan} G.~L.,   {Haiman} Z.,  2017, \mn@doi [\mnras]
  {10.1093/mnras/stx909}, \href
  {http://adsabs.harvard.edu/abs/2017MNRAS.469.1456V} {469, 1456}

\bibitem[\protect\citeauthoryear{{Weaver}, {McCray}, {Castor}, {Shapiro}  \&
  {Moore}}{{Weaver} et~al.}{1977}]{1977ApJ...218..377W}
{Weaver} R.,  {McCray} R.,  {Castor} J.,  {Shapiro} P.,   {Moore} R.,  1977,
  \mn@doi [\apj] {10.1086/155692}, \href
  {http://adsabs.harvard.edu/abs/1977ApJ...218..377W} {218, 377}

\bibitem[\protect\citeauthoryear{{Wentzel}}{{Wentzel}}{1968}]{1968ApJ...152..987W}
{Wentzel} D.~G.,  1968, \mn@doi [\apj] {10.1086/149611}, \href
  {http://adsabs.harvard.edu/abs/1968ApJ...152..987W} {152, 987}

\bibitem[\protect\citeauthoryear{{Wentzel}}{{Wentzel}}{1971}]{1971ApJ...163..503W}
{Wentzel} D.~G.,  1971, \mn@doi [\apj] {10.1086/150794}, \href
  {http://adsabs.harvard.edu/abs/1971ApJ...163..503W} {163, 503}

\makeatother
\end{thebibliography}
\appendix
\section{Star formation model with supernova feedback}
\label{sec_sfr}

In \citet{2014NewA...30...89S},
we discussed star formation models incorporating supernova feedback 
that are constrained by observations of UV luminosity
functions of Lyman-break galaxies in the redshift range $1.5 - 10$
and the correlations found between star formation rate (SFR),
stellar mass and gas phase metallicity.
Here, we broadly follow the same prescriptions with some modifications
that are needed in order to get more accurate time resolved star formation rate in the
low mass galaxies that are likely to be important in spreading metals
into the IGM.

Consider a dark matter halo of mass $M$ which virialises 
at redshift $z_c$ and
starts to accrete baryonic 
mass $M_g(t)$.
A total fraction, $f_*$, of the accreted gas 
is assumed to be in the cold phase
and available for star formation. 
At a given time a portion of 
this cold
gas is locked up
in low mass stars formed from previous episodes of star formation.
Further, massive stars
explode as supernova after a characteristic life time
of $t_{\rm SNe} \sim 10^7$~yrs, and  
inject energy and momentum to the surrounding gas.
A fraction of 
cold gas 
will then leave the galaxy making it 
unavailable for subsequent star formation.

Taking these processes in to account, and assuming that 
the instantaneous star formation rate, $d M_*(t)/dt$, 
of a galaxy is proportional to its cold gas content, we have
\begin{equation}
\frac{d M_*(t)}{dt} = f_t \left[\frac{ f_* M_g (t) - M_* (t) - M_w (t) }{ \tau }\right].
\label{eqn_dmdt}
\end{equation}
Here $\tau$ is the dynamical time scale for the halo and
$f_t$ governs the duration of star formation activity 
in terms of this time scale. 
Further $M_*(t)$ is the stellar mass at any time $t$,
$M_w(t)$ is the mass of cold gas lost in the wind
and the numerator of Eq.~\ref{eqn_dmdt} reflects the cold gas
mass still available for star formation.
The time
derivative of Eq.~\ref{eqn_dmdt} leads to
 \begin{equation}
\frac{d^2M_*(t)}{dt^2}=\frac{f_t}{\tau}\left[f_*\frac{dM_g(t)}{dt} - \frac{d M_*(t)}{dt} - \frac{dM_w(t)}{dt}\right]. 
\label{d2Mdt}
\end{equation}
In order to solve for the star formation rate one needs
to specify both baryonic accretion rate and the mass loss
rate of cold gas in the outflow.
 
We assume
that the baryon accretion rate in the central part of galaxy, 
a time $t$ after halo formation, goes as
\begin{equation}
\frac{dM_g}{dt}=\left(\frac{M_b}{\tau}\right)e^{-\frac{t}{\tau}}.
\label{eqn_accretion}
\end{equation}
Here $M_b$ is
the total baryonic mass that would be accreted in the halo and the
characteristics time scale for accretion 
is taken to be 
the dynamical time scale for the halo.

The mass outflow rate due to SNe explosion at time $t$ 
is assumed proportional to the star formation rate 
at an earlier time $(t-t_{\rm SNe})$ (to take account
of the delay between star formation and resulting supernovae), 
{\it i.e.}, 
\begin{equation}
\dot{M}_w (t) = \eta_w \dot{M}_{*}(t-t_{\rm SNe}).
\label{eqn_eta}
\end{equation}
where over dot represents the
time derivative. Depending on the actual mechanism of the outflow, the
proportionality constant $\eta_w$ is either $\propto v_c^{-2}$ (energy
driven/ cosmic ray driven outflows) or $ \propto v_c^{-1}$ (momentum
driven outflows) ; $v_c$ being the circular velocity of the dark matter
halo hosting the galaxy.
Further, we showed in \citet{2014NewA...30...89S}
that a $\eta_w \propto v_c^{-2}$ is
consistent with the amount of stellar mass detected in low mass dwarf galaxies.
We also show that in presence of cosmic rays,
the outflow never transits to a momentum driven stage.
Note that we will use the notation $\eta_w=(v_c/v_c^*)^{-\alpha}$ with $v_c^*$
fixing the normalisation and different values of $\alpha$ indicating
the type of outflow.

Using Eq.~\ref{eqn_accretion} and Eq.~\ref{eqn_eta} in Eq.~\ref{d2Mdt}
then gives
 \begin{equation}
\frac{d^2M_*(t)}{dt^2}=
\frac{f_t}{\tau}\left[f_*\frac{M_b}{\tau}e^{-\frac{t}{\tau}} - \frac{d M_*(t)}{dt} - \eta_w\frac{dM_*(t-t_{\rm SNe})}{dt}\right].
\label{eqn_dd}
\end{equation}
Note the explicit $t$ dependence here. 
This is only true for $t > t_{\rm SNe}$. 
For $t \le t_{\rm SNe}$ 
we have no supernova feedback and then,
 \begin{equation}
\frac{d^2M_*(t)}{dt^2}=\frac{f_t}{\tau}\left[f_*\frac{M_b}{\tau}e^{-\frac{t}{\tau}} - \frac{d M_*(t)}{dt} \right].
\label{eqn_d2M}
\end{equation}

In order to follow the star formation rate of individual galaxies
for $t > t_{\rm SNe}$,
we solve the delayed differential equation (Eq.~\ref{eqn_dd}) numerically.
For $t < t_{\rm SNe}$, one can integrate Eq.~\ref{eqn_d2M} analytically 
with the boundary condition that $dM_*/dt=0$
at $t=0$ and get
\begin{equation}
\frac{d M_*}{dt} = \frac{M_b f_* f_t}{\tau [f_t-1]}\Bigg[ e^{-\frac{t}{\tau}} - e^{-f_t\frac{t}{\tau}} \Bigg].
\label{eqn_dM_dt_2}
\end{equation}
Further, for simplicity we assume $f_t=1$ \citep[as in][]{2014NewA...30...89S}
and therefore
\bea
\dot{M}_{*}(t) &=& \frac {f_{*} M_b}{\tau} \frac{t}{\tau}  
\exp\left[-\f{t}{\tau}\right].
\label{eqnsf}
\eea
for $t < t_{\rm SNe}$. This evaluated at $t = t_{\rm SNe}$,
also provides the initial condition for the numerical solution of
Eq.~\ref{eqn_dd}.

\begin{figure}
\centerline{
\epsfig{figure=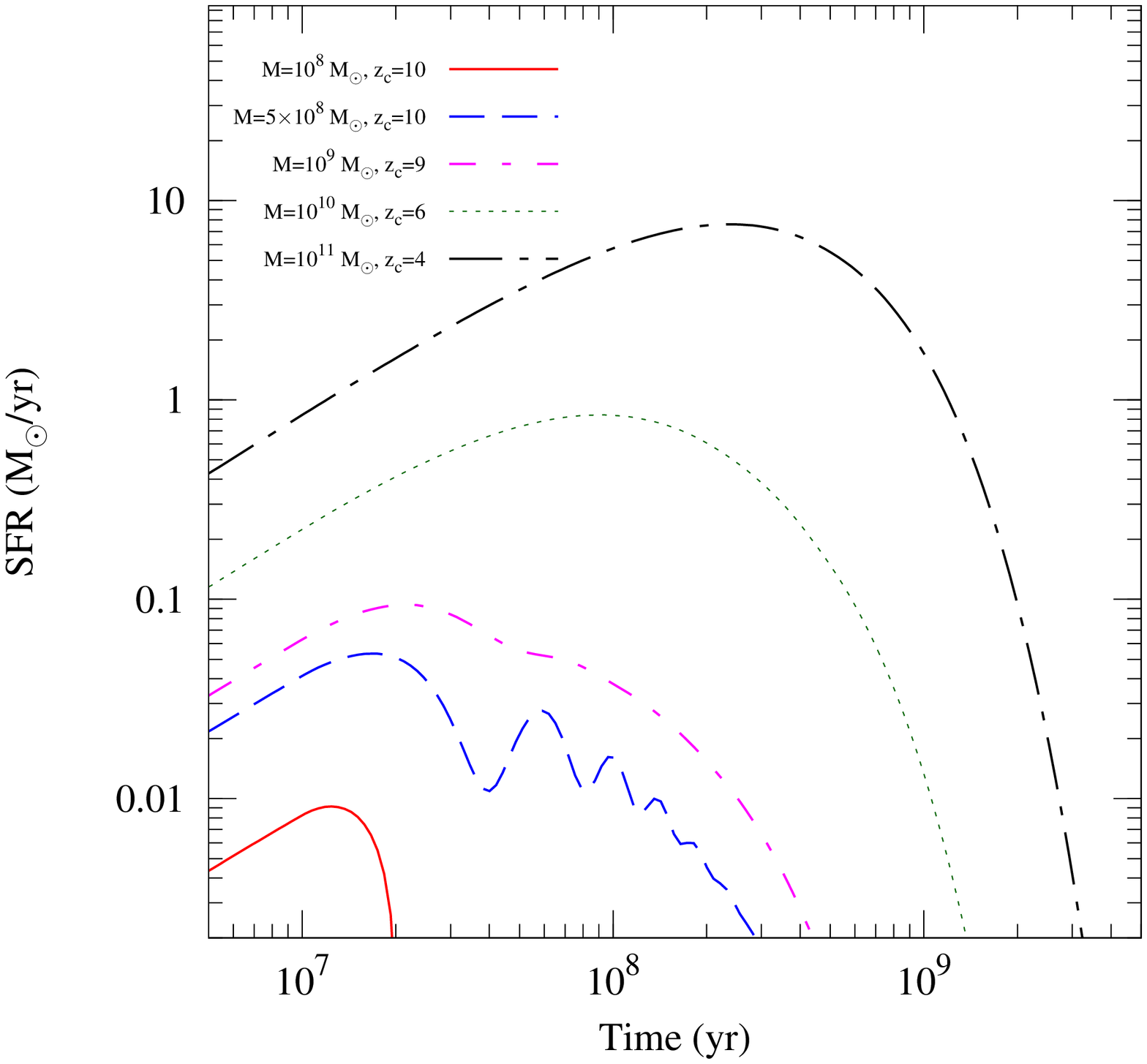,width=8.0cm,angle=-90.}}
\caption[]{Time evolution of SFR for different galaxies; The halo mass ($M$) and the collapsed
redshift ($z_c$) are marked in the legends. We have assumed $\alpha=2, ~ v_c^* = 100$~km/s.
}
\label{fig_sfr}
\end{figure}

In Fig.~\ref{fig_sfr} we show the resulting star formation rate of few example
galaxies with $\alpha=2$ and $v_c^*=100$~km/s. 
Further, we assume $f_*=0.5$. Note this $f_*$ is not what is
canonically referred 
to as the star formation efficiency in literature. Rather, the total 
fraction of baryons converted to stars in a galaxy over its lifetime
would be $f_*/(1+\eta_w)$.
These representative samples
are chosen to show a range in the time evolution of the star formation rate
that can be envisaged in our models.
For very low mass galaxies such as the one with $M=10^8~M_\odot$, we see a complete suppression
of star formation once the wind starts after $t_{\rm SNe}$ (solid red curve).
In these galaxies the mass outflowing rate or $\eta_w$ is very high. 
The star formation which happens within $t_{\rm SNe}$ is enough to
eject rest of the gas from
the shallow dark matter potential quenching further star formation completely.
 Galaxies with
slightly larger masses ($M=5\times 10^8~M_\odot$) show periodic/episodic
star formation activity (dashed blue curve). While $\eta_w$ is not high enough
to suppress the star formation completely, 
it definitely lowers the star formation that causes a reduction in subsequent
mass outflow as well. This in turns increases the gas reservoir and hence
the star formation, resulting a oscillating star formation mode in the galaxy.
Note that these two modes of star formation i.e. complete suppression and
episodic star formation would not have been seen in our model unless
we consider the delay of $t_{\rm SNe}$ between star formation and onset
of outflows.

Larger mass galaxies ($M=10^9~M_\odot$) do not show such oscillations
in star formation
but suppression due to outflows is clearly noticeable (magenta dotted-dashed
curve in Fig.~\ref{fig_sfr}). For very massive galaxies (i.e. 
$M\gtrsim 10^{11}~M_\odot$), there is negligible suppression of star formation.
Due to deeper gravitational potential, the mass outflow rate is small in those galaxies,
hence there is no strong suppression
of star formation due to the wind feedback. 
Note that we only consider $\alpha=2$ as we will show
later that the outflows would never be going to a momentum driven
case in presence of cosmic rays. Also the normalisation
circular velocity $v_c^*=100$~km/s is chosen because it reproduces various
observations regarding high redshift universe \citep[see][]{2014NewA...30...89S}.
These star formation models are used to calculate the outflow
properties in the main text.

\section{Shock jump conditions in presence of cosmic rays}
\label{appendix1}
The shock jump conditions of two fluid models with one relativistic
component ($\gamma = 4/3$) and one non-relativistic component
($\gamma = 5/3$) were evaluated
by \citet{1983ApJ...272..765C}.
They are
\citep[Eqn.~5 to 8 of][]{1983ApJ...272..765C},
\begin{eqnarray}
\rho_2 &=& \dfrac{\gamma_s + 1}{\gamma_s - 1} \rho_1 \\
v_2 &=& \dot{R}_1 +  \dfrac{\gamma_s - 1}{\gamma_s + 1} \left( v_w - \dot{R}_1\right) \\
P_b &=& \dfrac{2(1-w)}{\gamma_s+1} \rho_1 \left(\dot{R}_1 - v_w\right)^2 \label{eqn_PbA} \\
P_c &=& \dfrac{2w}{\gamma_s+1} \rho_1 \left(\dot{R}_1 - v_w\right)^2 \label{eqn_PcA}\\
\gamma_s &=& \dfrac{5+3w}{3(1+w)},
\end{eqnarray}
 with $w=P_c/(P_c+P_b)$. Here, subscript 1 and 2 are for pre-shocked and
post-shocked properties respectively. Further, we assume that $w$ remains
constant across the shock.
Adding, Eqn.~\ref{eqn_PbA} and Eqn.~\ref{eqn_PcA}
we obtain Eq.~\ref{R1eqn}. Putting $w$ explicitly in terms of $P_c$ and $P_b$
one can get Eq.~\ref{gammas_eqn}.  

\bsp	
\label{lastpage}

 \end{document}